\documentclass[12pt]{article}

\usepackage{epsfig}
\usepackage{amsfonts}
\usepackage{amsopn}
\usepackage{amsmath}
\usepackage{verbatim,amsthm}

\title
{
\vskip-50 pt
\begin{flushright}
\normalsize\rm NORDITA-2015-98
\end{flushright}
\vskip 20 pt
 Gauge theory approach to branes and spontaneous symmetry breaking
}
\author{
 A. A. Zheltukhin $^{a,b}$\thanks{e-mail: aaz@physto.se
 } 
  \\ \\
$^a$ Kharkov Institute of Physics and Technology, 
Kharkov, 61108, Ukraine \\  
$^b$ Nordita, KTH Royal Institute of Technology and Stockholm University\\
SE 106 91 Stockholm, Sweden
}

\begin{document}
\maketitle
\begin{abstract}
 
 Gauge theory approach to consideration of the Nambu-Goldstone  bosons as 
 gauge and vector fields represented by the Cartan forms of spontaneously 
 broken symmetries, is discussed. The approach is generalized to describe the 
 fundamental branes in terms of $(p+1)$-dimensional  worldvolume gauge and massless 
 tensor fields consisting of the Nambu-Goldstone bosons associated with 
 the spontaneously broken Poincare symmetry of the $D$-dimensional Minkowski space.

\end{abstract}

\section{Introduction}

Equivalence between explicit solution of the string equations 
and restoration of its worldsheet in the Minkowski space was observed 
by Regge and Lund \cite{RL} (see also [2-4]\nocite{Om, BNCh, BN}). 
This statement resulted from the differential geometry of embedded 
surfaces, in particular, from the first-order linear  Gauss-Weingarten 
 differential equations 
for the vectors tangent and normal  to string worldsheet \cite{Eisn}. 
 In the case of 4-dim. Minkowski space they found that the integrability 
 condition for the G-W equations was just the sine-Gordon equation describing 
 a surface embedded into 3-dimensional Euclidean space. 
 This connection of the nonlinear string equations with the 
 linear equations of the inverse scattering method \cite{AKNS}
 added a new sight to numerous efforts to overcome nonlinearities of the (mem)brane 
 dynamics [7-19] \nocite{ hoppe1, BST, DHIS, Nic,
 WHN, FI, WLN,  BZ_0, Twn, Wit,  Duff, TZ, Hop}. 

 The geometric approach by Regge-Lund was generalized to strings embedded 
 into $D$-dim. Minkowskii space in \cite{Zgau2}, \cite{Zgau4}, 
 where its reformulation in terms of the Yang-Mills theory was developed. 
 The generalization was based on the Cartan method of moving frame  \cite{Car}
and its development by Volkov \cite{Vol1} under geometrization of 
 the phenomenological Lagrangian method [24-27] \nocite{Wei, Schw, CWZ, CCWZ}.
 The gauge reformulation established equivalence between 
 the string and a closed sector of states of the exactly 
 integrable 2-dim. $SO(1,1)\times SO(D-2)$ 
 invariant model of interacting gauge and massless scalar fields.
This approach was recently generalized to the case of the fundamental 
Dirac $p$-branes embedded into $D$-dim. Minkowski space, 
where the $p$-brane turned out to be the exact solution of 
a $(p+1)$-dimensional model invariant under diffeomorphisms 
and  $SO(D-p-1)$ gauge transformations \cite{Zbran}. 
The model contains the constrained $SO(D-p-1)$ multiplets 
 of gauge and massless tensor fields. The latter represent the 
second fundamental form  of $(p+1)$-dim. world hypervolume.
 This reformulation is based on the construction by Faddeev and 
Semenov-Tyan-Shansky showing the equivalence 
between chiral and constrained Yang-Mills field theories \cite{FST}. 
 
Here we develop this approach as an alternative
way to describe the Nambu-Goldstone (N-G) fields of a spontaneously 
broken internal symmetry $G$ in terms of the constrained vector 
and Yang-Mills multiplets of the unbroken subgroup $H \in G$. 
Such a description treats the multiplet components as new 
dynamical variables 
of an associated gauge invariant action.
The gauge covariant constraints are the integrability
 conditions of the PDEs which express the N-G fields through 
the multiplet fields. 
These conditions are just the Maurer-Cartan (M-C) equations 
for the space  $G/H$ represented in terms of the gauge strength 
and vector fields.
We explain how the constraints reduce the redundant components of 
the Cartan multiplets preserving the number of the physical DOF equal to  
the number of the essential N-G fields. Using these results we 
 prove that the multiplets of the  p-brane model \cite{Zbran} 
represent the N-G fields of the broken  $ISO(1, D-1)$ global symmetry 
of $\mathbf{R}^{1,D-1}$. 
The Poincare symmetry breakdown is caused by the embedding of 
a $(p+1)$-dimensional hypersurface $\Sigma_{p+1}$ 
into $\mathbf{R}^{1,D-1}$.
 We find that the brane vacuum manifold is fixed by the $(D-p-1)$
conditions: $\omega_{a}\equiv\mathbf{n}_{a}(\xi)d\mathbf{x}(\xi)=0$ 
for the local translations orthogonal to $\Sigma_{p+1}$ at each of its
 points $\mathbf{x}(\xi)$ \cite{Zgau2}. 
These conditions are invariant under the {\it left global} $ISO(1, D-1)$  
and the {\it right local} $ISO(1, p)\times SO( D-p-1)$ transformations 
 of the Cartan moving frame $\mathbf{n}_{A}(\xi)$ attached to $\Sigma_{p+1}$
 and its origin $\mathbf{x}(\xi)$. 
The world vectors $\mathbf{x}(\xi)$ 
and $\mathbf{n}_{A}(\xi)$ are treated as macroscopic translational 
and rotational order parameters, and $\Sigma_{p+1}$ emerges as 
a world hypervolume swept by the Dirac $p$-brane. 
It follows from the  $SO(D-p-1)$ gauge invariant 
action $S_{Dir}$ (\ref{actnl}) formulated in terms of 
the Cartan multiplets located on the minimal hypersurface $\Sigma_{p+1}$.  
This action is also invariant under diffeomorphisms because the gauge 
fields of the unbroken gauge subgroup $SO(1, p)$ are identified with 
the metric connection of $\Sigma_{p+1}$. The effective tensor 
multiplet $l_{\mu\nu}{}^{a}(\xi)$ encoding the broken rotational 
and translational  N-G modes turns out to be the traceless 
second  quadratic form of $\Sigma_{p+1}$. 
	The modes associated with the translations in $\mathbf{R}^{1,D-1}$ 
are also presented in $S_{Dir}$ by non-dynamical  
components of the background metric $g_{\mu\nu}(\xi)$ of $\Sigma_{p+1}$. 
These modes provide invariance of $S_{Dir}$ under diffemorphisms and generate
 the cosmological term $\sqrt{|g|}$ which is dynamical one in the standard 
  Dirac-Nambu action. But  in the gauge approach the 
 dynamical equations for $g_{\mu\nu}$ are provided by the Gauss conditions.
 The latter, together with the Ricci embedding conditions, are 
 taken into account in the Euler-Lagrange EOM following from $S_{Dir}$. 
 For the case of string in 3-dim.
 Minkowski space the Gauss constraints turn out to be equivalent to 
 the condition $R_{\mu\nu}=\kappa g_{\mu\nu}$ defining Einstein spaces 
 \cite{Zgau2}. 
We find that the G-C-R constraints considered as the initial 
data of the Cauchy problem for the brane EOM are conserved.
It proves the uniqueness of the found solution of the Euler-Lagrange 
PDEs as a consequence of the Cauchy-Kowalevskaya theorem. 
 
The nonlinear realization of the Poincare group broken to the discussed subgroup
was used in many papers for description of generalized string and $p$-brane dynamics
 (see e.g. [30-36] \nocite{BCGM, GKW, CLNVX, GM, AKo, GKP, BZ_hedr} 
 and refs. there).
Construction of effective string and brane actions including 
  higher-derivative terms in the world vector $\mathbf{x}(\xi)$ 
  was studied there. 
  Various methods, including the use of the Cartan forms,  
were proposed to build the Lorentz and diffeomorphism invariant 
higher order terms. However, any systematic scheme producing such invariants  
  faces the problem of their classification in each order
 in derivatives,  because not all admissible invariants are independent. 
The Cartan form products together with their covariant exterior differentials 
 make up a complete set of invariants contributing to the effective action. 
 This observation was earlier taken into account in our paper \cite{VGZP},
 where a general method was proposed for constructing the individual terms 
of the expansion of the $n$-point dual amplitude with respect to homogeneous
 functions of degree $r=1,2,...$ of the Mandelstam kinematic invariants $s_{ik}$. 
The fulfillment of the Adler's principle  
was ensured there by using a phenomenological Lagrangian invariant under
 the chiral symmetry and containing higher derivatives of the meson fields. 
The complete set of the invariants of the fourth order in the derivatives was  
built in \cite{VGZP} for an arbitrary semisimple symmetry group realized non-linearly.
These results are used in the considered here gauge approach to branes.
The Cartan multiplets and their covariant derivatives create covariant blocks for 
constructing higher order terms in effective action of strings and $p$-branes 
 explicitly invariant under the Lorentz transformations and diffeomorphisms. 
This scheme is considered  in Sections 2,3, where the general method of phenomenological
 Lagrangians is shortly dwelt on.
Application of the discussed method to broken Poincare symmetry
 yields the desired higher order invariants.
For example, the invariant terms in the gauge invariant actions for the  Nambu-Goto 
string \cite{Zgau2}, \cite{Zgau4} and Dirac $p$-brane \cite{Zbran} are quadratic
 in covariant derivatives of the second fundamental form and quartic in its components.
 In view of the generalized $\it {Gauss \, Theorema  \, Egregium}$  \cite{Zbran} the quartic 
terms are proportional to the ones quadratic in the Riemannian tensor components.
 Similar result was observed  in \cite{GM}, where for the first leading correction to the
 Nambu-Goto string action was found to be quadratic in the world sheet curvature.
 So, unification of the Cartan approach explicitly covariant under the Lorentz transformations 
and diffeomorphisms with the methods \cite{GM}, \cite{AKo} 
is promising and deserves a special investigation.
It seems also interesting to apply the gauge approach to the 
quantization problem of Dirac branes. 
The point is that the gauge approach yields
a map of the brane dynamics in the dynamics of a Yang-Mills model which obeys the 
universal embedding constraints for the world hypersurfaces swept by branes. 
The Cartan multiplets represent the covariant objects realizing the said map. 
 The gauge invariant $p$-brane action $S_{Dir}$ is suitable  
 for application of the BRST-BFV method which has demonstrated its power in quantization of the  
Yang-Mills theories with constraints (see e.g. \cite{deW}, \cite{FV}). 
This method will yield an important information about the ghost sector structure, anomalies and 
 critical dimension in the theory of branes.
 The use of the gauge approach also gives a clear geometric explanation 
for many aspects of the  classical brane dynamics reformulated 
in terms of the Cartan %moving polyhedron$(\mathbf{x}(\xi),\mathbf{n}_{A}(\xi))$. 
multipletd. So, the inverse Higgs phenomenon \cite{IO} which pulls the trigger for 
 splitting  N-G modes into essential and not essential ones means a simple  
choice  for the first $(p+1)$ vectors $\mathbf{n}_{i}(\xi)$ of the moving 
frame $\mathbf{n}_{A}(\xi)$ to be tangent to $\Sigma_{p+1}$. 
It is equivalent to the above mentioned definition of the vacuum manifold by 
the conditions $\omega_{a}=0$. 
Another example of the inverse Higgs conditions emerging  in the gauge theory 
approach to branes is given by the minimality conditions 
$Spl^{a}\equiv g^{\mu\nu}l^{a}_{\mu\nu}=0$ 
invariant under all symmetries of the action $S_{Dir}$.

\section {Coset spaces and fiber bundles}

Consider the Nambu-Goldstone fields generated by a global semisimple group
  of symmetry $G$ with the algebra generators $X_{i}$ and $Y_{\alpha}$
    \begin{eqnarray}\label{comr}
[Y_{\alpha}, Y_{\beta}]= ic^{\gamma}{}_{\alpha\beta}Y_{\gamma}, \ \
[X_{i}, Y_{\alpha}]= ic^{k}{}_{i\alpha}X_{k}, \ \
[X_{i}, X_{k}]= ic^{\alpha}{}_{ik}Y_{\alpha} + ic^{l}{}_{ik}X_{l}.
\end{eqnarray}
If  $G$ is a spontaneously broken symmetry of a physical system with its vacuum invariant 
under the subgroup $H$, it is convenient to describe the corresponding N-G fields 
(in some neighborhood of the identity of $G$)
 with the help of the factorized representation of the group elements 
(\cite{Vol1},\cite{CWZ},\cite{CCWZ}) 
\begin{eqnarray}\label{fctr} 
G(a,b)= K(a)H(b),
\end{eqnarray}
where $a$ and $b$ parametrize the group space of $G$.
 The left multiplication of the  global group $G$ by any element $g \in G$ 
 \begin{eqnarray}\label{actn}
 gG=G'  \ \ \    \rightarrow  \ \ \ \     gK(a)H(b)=K(a')H(b')  
 \end{eqnarray}
 yields the following  transformation rules of the parameters $a$ and $b$
\begin{eqnarray}\label{trf} 
a'=a'(a,g),  \ \ \ \ b'=b'(b,a,g),
\end{eqnarray}
with the transformed parameters $a'$ independent of $b$. 
This shows that the parameters $a$ form the left coset space $G/H$ 
invariant under non-linear transformations of $G$. Then these parameters  
 may be mapped into the components of the N-G field $\pi(x)$ with the same 
nonlinear transformation law (\ref{trf})
\begin{eqnarray}\label{trfNG} 
\pi'(x)= a'(\pi(x),g).
\end{eqnarray}
One can see that  the left multiplication (\ref{actn}) preserves the form $G^{-1}dG$ 
\begin{eqnarray}\label{lshi}
  G'^{-1}dG'=G^{-1}dG.
\end{eqnarray}
As a result, the expansion of the one-form (\ref{lshi}) in the generators $X_{i}, Y_{\alpha}$ 
\begin{eqnarray}\label{cafo}
 G^{-1}dG= i\omega_{G}^{i}(a,b,da)X_{i} + i\theta_{G}^{\alpha}(a,b,da,db)Y_{\alpha},
\end{eqnarray}
taking into account their algebra (\ref{comr}), creates  the  Cartan differential 
 one-forms $\omega_{G}^{i}$ and $\theta_{G}^{\alpha}$ \cite{Car} invariant under 
 the left multiplications (\ref{actn})
\begin{eqnarray}\label{Gfrm}
\omega_{G}^{i}(a',b',da')=\omega_{G}^{i}(a,b,da), \ \ 
\theta_{G}^{\alpha}(a',b',da',db')=\theta_{G}^{\alpha}(a,b,da,db). 
\end{eqnarray}
Therefore, it was proposed to use these forms as building blocks 
 for construction of $G$-invariant phenomenological Lagrangians \cite{Vol1}.

The parameters  $b$ in the forms (\ref{cafo}) describe non-physical DOF 
and can be fixed like gauge parameters. 
This becomes clear after the substitution of expression (\ref{fctr}) into  $G^{-1}dG$ 
that yields the new representation for this form
\begin{eqnarray}\label{carfo} 
G^{-1}dG= H^{-1}(K^{-1}dK)H + H^{-1}dH 
\end{eqnarray}
which shows that the $b$-dependence of $G^{-1}dG$ arises as a result of $H$-subgroup 
gauge transformations of the $b$-independent form $K^{-1}dK$.

After taking into account Eq. (\ref{carfo}), condition (\ref{lshi}) transforms into 
\begin{eqnarray}\label{lshi1}
H'^{-1}(K'^{-1}dK')H' + H'^{-1}dH'= H^{-1}(K^{-1}dK)H + H^{-1}dH,
\end{eqnarray}
where $K'H'=gKH, \ \    K'\equiv K(a'(a,g)),  \ \  H'\equiv H(b'(a,b,g))$.
 
Condition (\ref{lshi1}) may be rewritten in the form of the transformation law 
\begin{eqnarray}\label{lshi2}
K'^{-1}dK'=\tilde{H}(K^{-1}dK)\tilde{H}^{-1} + \tilde{H}d\tilde{H}^{-1},  
 \ \ \  \tilde{H}:=H'H^{-1}
\end{eqnarray}
that  shows that the reduced Cartan form $K^{-1}dK$, depending only on the coset parameters $a$ 
and their differentials
\begin{eqnarray}\label{cosfrm}
K^{-1}dK= i\omega^{i}(a,da)X_{i} + i\theta^{\alpha}(a,da)Y_{\alpha},
\end{eqnarray} 
transforms like the one-form of the gauge potential for $H$. 
The forms $\omega^{i}(a,da)$ and $\theta^{\alpha}(a,da)$ are the forms (\ref{Gfrm}) 
calculated at the point $a$ in the fixed coordinate frame with $b=0$. 
Therefore, these forms are not  invariant under the global $G$-shifts and  their 
 transformation rules 
\begin{eqnarray}\label{frmtr}
\omega'^{i}X_{i} =\omega^{i}(\tilde{H}X_{i}\tilde{H}^{-1}), \ \ \ \ \
\theta'^{\alpha}Y_{\alpha}= \theta^{\alpha}(\tilde{H}Y_{\alpha}\tilde{H}^{-1}) - i\tilde{H}d\tilde{H}^{-1}
\end{eqnarray}
follow from Eq. (\ref{lshi2}) after using the commutation relations (\ref{comr}). 

For infinitesimal transformations $g\in G$ in the fixed basis $b=0$ one can write $b'$ and 
$\tilde{H}$ in the following form 
\begin{eqnarray}\label{inftr}
	b'= \epsilon(a,0,g),  \ \ \ \ \ \   \tilde{H}\approx 1 + i\epsilon^{\beta}(a,0,g)Y_{\beta}.
	\end{eqnarray}
 As a result, Eqs. (\ref{frmtr})  are presented in the matrix form 
 \begin{eqnarray}\label{frmtr'}
\delta\omega^{i}X_{i} =i\epsilon^{\beta}[Y_{\beta}, X_{i}]\omega^{i},
 \ \ \ \ \
\delta\theta^{\alpha}Y_{\alpha}= i\epsilon^{\beta}[Y_{\beta}, Y_{\alpha}]\theta^{\alpha}
 -d\epsilon^{\gamma} Y_{\gamma}.
\end{eqnarray}
which realizes the adjoint representation of the vacuum subgroup $H$   
\begin{eqnarray}\label{inftrf}
\delta\omega^{i} =c^{i}{}_{k\beta}\epsilon^{\beta}\omega^{k},  \ \ \ \ \ \ \
\delta\theta^{\alpha}= -d\epsilon^{\alpha} + c^{\alpha}{}_{\gamma\beta}\epsilon^{\beta}\theta^{\gamma}.
\end{eqnarray}
 with the infinitesimal  parameters $\epsilon^{\alpha}$ depending on the coset coordinates $a$.

This shows that various combinations of the covariant Cartan forms $\omega^{i}$ (\ref{cosfrm})
 invariant under the gauge group $H$ are automatically invariant under the left global transformations 
of $G$, and can be used for construction of the action of the N-G bosons \cite{Vol1}. 
The Cartan forms $\theta ^{\alpha}$ are the gauge potentials that permit to introduce covariant 
differentials for any multiplet of $H$. 
Thus, $\theta ^{\alpha}$ may be used not only for description of interactions between the 
N-G fields, but also of their interactions with other fields.

 So, we have called to mind the well-known method of non-linear realizations using N-G fields 
associated with the spontaneously broken symmetry $G$. The method presents  the  
parameter space of $G$ as a fiber bundle where the coset $G/H$ is the base 
and $H$ is the fiber. A special accent has been made on some properties of the 
reduced Cartan one-forms $\omega^{i}, \, \theta^{\alpha}$ (\ref{cosfrm}), because  
they realize linear representations, and can be alternatively used as new 
effective field variables instead of the usual coset 
coordinates $a^{\lambda}$. Due to  the Gauss-Codazzi conditions for the Cartan forms
 and their gauge covariance, they do not produce new physical degrees of freedom, but only
 encode the N-G fields in terms of the constrained gauge multiplets of the vacuum 
subgroup $H$ \cite{FST}. This approach to the description of N-G fields does not use
 any explicit parametrization of G and reveals their pure geometric roots based 
on the conception of connections and gauge fields. 
In this geometrical description of  N-G fields their  Euler-Lagrange equations,  
derived from a given phenomenological action, take the 
form of some gauge covariant constraints for the Cartan multiplets.
To find the form of these constraints we need at least to choose an invariant 
phenomenological Lagrangian for the N-G fields. It will be considered 
in the next Section  where some known examples of invariant
 Lagrangians  formulated in terms of the Cartan forms will be given.

\section {Phenomenological Lagrangians}

Going from the coset parameters $a$ to the N-G fields $\pi(x)$ and using 
the definition of the differential one-forms $\omega^{i}$ 
\footnote{The greek letters $\lambda, \mu, ...$ from the second half of the
 alphabet numerate the coordinate indices of the curved space $G/H$.}
\begin{eqnarray}\label{omeg}
\omega^{i}(\pi, d\pi)=\omega^{i}_{\lambda}(\pi)d\pi^{\lambda}
\end{eqnarray}
 one can see that the Lame coefficients  $\omega^{i}_{\lambda}(\pi)$  
have the meaning of the vielbein components of the curved space $G/H$, that permits to write
 the squared element of length in the tangent space to $G/H$ as 
\begin{eqnarray}\label{mtr}
ds^{2}=\omega_{i}\omega^{i}
=\omega^{i}_{\lambda}\omega_{i \rho}d\pi^{\lambda}d\pi^{\rho}\equiv g_{\lambda\rho}(\pi)d\pi^{\lambda}d\pi^{\rho}, 
\end{eqnarray}
where the invariant non-degenerate Cartan-Killing tensor built from 
the structure constants of the algebra 
(\ref{comr}) is used as a metric tensor to raise the subindex $i$ of $\omega_{i}$.
The representation  $g_{\lambda\rho}(\pi)=\omega^{i}_{\lambda}\omega_{i\rho}$ used in (\ref{mtr})
 is accompanied with the completeness condition 
\begin{eqnarray}\label{compl}
\omega^{i}_{\lambda}(\pi)\omega^{\lambda}_{j}(\pi)=\delta^{i}_{j}
\end{eqnarray}
that permits  to invert relation (\ref{omeg}) expressing  $d\pi^{\lambda}$ in terms of $\omega^{i}(d)$
\begin{eqnarray}\label{invers}
d\pi^{\lambda}=\omega^{\lambda}_{i}\omega^{i}(d),
\end{eqnarray}
and the space-time derivative $\partial_{m}\pi^{\lambda}\equiv \frac{\partial\pi^{\lambda}}{\partial x^{m}}$ 
 through the  vector field $\omega^{i}_{m}
 $\footnote{The latin letters $m,n,p, ...$ placed after the letter $l$ in the alphabet 
 are used for the space-time coordinates $x^m$.}
\begin{eqnarray}\label{invers'}
\partial_{m}\pi^{\lambda}%\equiv \pi^{\lambda}_{,m}(x)
=\omega^{\lambda}_{i}(\pi(x))\omega^{i}_{m}(\pi(x)), \ \  \omega^{i}_{m}: =\frac{\omega^{i}(d)}{dx^m} \,.
\end{eqnarray}
The relations show that the transition to the vector 
fields $\omega^{i}_{m}(\pi(x))$ from $\omega^{i}_{\lambda}(\pi(x))$ is realized by projecting them on the 
derivatives $\partial_{m}\pi^{\lambda}$,  and this does not increase the DOF number. 
The simplest invariant action in the long-wave approximation for N-G fields
expressed in terms of $\omega^{i}_{m}(\pi(x))$ is \cite{Vol1}
\begin{eqnarray}\label{int}
S= \frac{\gamma}{2}\int d^{D}x \, \omega^{i}_{m}\omega^{m}_{i}  
\end{eqnarray}
and coincides with the action of massless scalar particles quadratic in $\partial_{m}\pi^{\lambda}(x)$
  in the curved space $G/H$, but  with its coordinates $\pi^{\lambda}$ parametrized by $x^{m}$
\begin{eqnarray}\label{intg}
S= \frac{\gamma}{2}\int d^{D}x \,g_{\lambda\rho}(\pi) \partial_{m}\pi^{\lambda}\partial^{m}\pi^{\rho}.
\end{eqnarray}
 The corresponding EOM for the field $\pi^{\lambda}(x^{m})$ is the  geodesic equation  
%following from $S$ (\ref{intg} is the equation 
\begin{eqnarray}\label{geod}
\partial_{m}\partial^{m}\pi^{\rho} + \Gamma^{\rho}_{\lambda\sigma}\partial_{m}\pi^{\lambda}\partial^{m}\pi^{\sigma}=0
\end{eqnarray}
in the Riemannian space $G/H$  with its Christoffel symbols $\Gamma^{\rho}_{\lambda\sigma}(\pi^{\sigma}(x))$.

Another example of an invariant action is provided  by using the gauge 
form $\theta^{\alpha}$ for $H$ that  follows from (\ref{inftrf}). This form permits to construct  
the gauge covariant exterior differential $D\wedge\omega^{i}$ %for the form $\omega^{i}$  
\footnote{The symbols $\wedge$ and $d\wedge$ denote the exterior product and the external 
differential of the differential forms, respectively.}
\begin{eqnarray}\label{covd}
D\wedge\omega^{i} 
=d\wedge\omega^{i} + c^{i}{}_{k \alpha}\theta^{\alpha}\wedge \omega^{k}. 
\end{eqnarray}
 with a homogeneous gauge transformation under multiplications  (\ref{actn}) 
\begin{eqnarray}\label{codif}
\delta (D\wedge\omega^{i}) = c^{i}{}_{k\beta}\epsilon^{\beta} (D\wedge\omega^{k}),
\end{eqnarray}
as it follows from the Jacobi identity produced by the cyclic
 sum  of the commutators  $[[Y_{\alpha}, Y_{\beta}],X_{i}] + ...=0$ 
\begin{eqnarray}\label{Jab}
c^{l}{}_{\beta k}c^{k}{}_{\alpha i} + c^{l}{}_{\alpha k}c^{k}{}_{i \beta} 
+ c^{l}{}_{i \rho}c^{\rho}{}_{\beta\alpha}=0.
\end{eqnarray}
As a result, we obtain the
%Then the covariant exterior form (\ref{covd}) can be used  to construct 
 invariant action in four-dimensional space-time quartic in the derivatives of N-G fields 
\begin{eqnarray}\label{4actn}
S= \frac{\tilde\gamma}{2}\int d^{4}x \,D\wedge\omega_{i}\wedge D\wedge\omega^{i}. 
\end{eqnarray}
However, this action turns out to be equal to the well-known action
\begin{eqnarray}\label{4actn'}
S= \frac{\tilde\gamma}{8}c_{jk}{}^{f}c_{fil}\int d^{4}x \,\omega^{j}\wedge\omega^{k}\wedge\omega^{i}\wedge\omega^{l}
\end{eqnarray}
   proportional to the squared constant torsion of $G/H$,  
 as it follows from the group structure Eq. (\ref{Tor'}) discussed in the next section 
 \footnote{The physical role of the affine connections corresponding to the constant torsions
  for the chiral groups $G \times G$ was cleared up in \cite{VZT}.} 
 Thus, this action  belongs to the above-discussed set of invariant actions polynomial in $\omega^{i}$.

More general invariant actions in the $D$-dimensional Minkowski 
space can be built by using the covariant derivatives $D_{n}\omega^{i}_{m}$ 
\begin{eqnarray}\label{covder}
 D_{n}\omega_{m}^{i}:=\partial_{n}\omega_{m}^{i} +  c^{i}{}_{k \alpha}\theta_{n}^{\alpha}\omega_{m}^{k}
\end{eqnarray}
instead of the exterior differentials (\ref{covd}) connected  with  $D_{n}\omega^{i}_{m}$ 
by the relation 
\begin{eqnarray}\label{covdc}
D\wedge\omega^{i}= D_{\nu}\omega_{\mu}^{i}d\pi^{\nu}\wedge d\pi^{\mu}.
\end{eqnarray}
Using the relation $\omega^{i}(d)= \omega_{\mu}^{i}d\pi^{\mu}=\omega_{m}^{i}dx^{m}$
one can rewrite (\ref{covdc}) as 
\begin{eqnarray}\label{covdst}
D\wedge\omega^{i}= D_{n}\omega_{m}^{i}dx^{n}\wedge dx^{m}.
\end{eqnarray}
The covariant derivatives $D_{n}\omega^{i}_{m}$ 
form linear representations of the gauge 
group $H$ and the Lorentz group in $D$-dimensional Minkowski space. 
So, invariant combinations of $D_{n}\omega^{i}_{m}$, 
such as  $D_{n}\omega^{n}_{i}D_{m}\omega^{mi}$ or  
$D_{n}\omega^{mi}D^{n}\omega_{mi}$ as well as their 
higher monomials, are invariants 
of the left shifts of $G$.
 These combinations form a subset of building blocks composing 
a general invariant phenomenological Lagrangian.

At last, we have  the covariant two-form $F^{\alpha}$ of the gauge strength for $\theta^{\gamma}$ 
\begin{eqnarray}\label{strh}
F^{\alpha}:= D\wedge\theta^{\alpha} 
=d\wedge\theta^{\alpha} + \frac{1}{2}c^{\alpha}{}_{\gamma\beta}\theta^{\beta} \wedge \theta^{\gamma}.
\end{eqnarray}
It also undergoes homogeneous gauge transformations under $H$ induced by the left shifts (\ref{actn})
\begin{eqnarray}\label{strhl}
\delta F^{\alpha}= c^{\alpha}{}_{\gamma\beta}\epsilon^{\beta}F^{\gamma},
\end{eqnarray}
as it follows from the Jacoby identity for the  structure constants of $H$
$$
c^{\gamma}{}_{\alpha\bullet}c^{\bullet}{}_{\nu\rho} + c^{\gamma}{}_{\nu\bullet}c^{\bullet}{}_{\rho\alpha} 
+ c^{\gamma}{}_{\rho\bullet}c^{\bullet}{}_{\alpha\nu}=0.
$$
For a semisimple Lie algebra one can choose a basis in which the structure 
constants with all lower or upper indices are completely antisymmetric.
As a result, the combinations of the 2-form gauge strength $F^{\gamma}$ invariant 
under the subgroup $H$ are invariant under nonlinear transformations 
of the group $G$, and may also be used for 
 construction of  $G$-invariant actions including higher order 
 terms in the derivatives of $\pi^\lambda$. 
The kinetic term for the gauge field $\theta^{\alpha}_{m}(\pi)$ 
composed from the N-G fields
\begin{eqnarray}\label{gkin}
S= -\frac{\tilde\gamma}{4}\int d^{D}x \,  F^{\alpha}_{mn} F_{\alpha}^{mn}
\end{eqnarray}
 is an example of the Lagrangian of the fourth order in the derivatives of  $\pi^\lambda$ 
invariant under the group $G$. There are other important invariants.

So, the Cartan forms make up a complete set of covariant blocks for construction 
of general phenomenological action. However, not all monomials constructed from the  Cartan forms 
are linearly independent. he Cartan structure equations play an important 
role in search for the number of independent invariants.
 
This short survey of the well-known description of N-G fields allows to consider 
their above-mentioned  description in terms of new effective dynamical variables: the 
fields $\omega_{m}^{i}, \, \theta^{\alpha}_{m}$ forming the linear Cartan multiplets.

\section {N-G fields as Cartan multiplets}

A transition from some  dynamical variables to other ones implies  
 the choice of transition functions connecting these two sets. 
The transition has to preserve the number of the original physical degrees of freedom.  
When the number of new variables is larger than  the number of the original variables, then 
the  corresponding constraints and/or a gauge symmetry have to be added to reduce the abundance.
 The latter case is realized by the transition from the N-G fields $\pi^{\lambda}(x)$
to the massless composite vector $\omega_{m}^{i}(\pi(x))$  and the 
gauge fields  $\theta^{\alpha}_{m}(\pi(x))$  forming the Cartan multiplets. 
Indeed, the definition  (\ref{cosfrm}) of the $\omega^{i}$ 
and $ \theta^{\alpha}$ differential forms can be considered to be 
the matrix system of PDEs
\begin{eqnarray}\label{PDE}
dK= i\omega^{i}KX_{i} + i\theta^{\alpha}KY_{\alpha},
\end{eqnarray}
for the matrix $K(\pi(x))$ as a function of the given coefficients 
$\omega_{m}^{i}(\pi(x))$ and $\theta^{\alpha}_{m}(\pi(x))$. 
Eqs. (\ref{PDE}) express the total differential $dK$
through the products of $K$ with the Cartan forms. As a result, the N-G fields are 
presented as implicit functions of the massless multiplets. 
Solving  PDEs (\ref{PDE}) one can restore the original N-G fields encoded by the Cartan multiplets. 
The PDEs are rather nontrivial and put severe constraints on the Cartan multiplets. 
The constraints follow from  the integrability conditions of Eqs. (\ref{PDE})
\begin{eqnarray}\label{ic}
%d\wedge (K^{-1}dK)
 d\wedge(\omega^{i}(\pi,d\pi)K)X_{i} + d\wedge(\theta^{\alpha}(\pi,d\pi)K)Y_{\alpha}=0
\end{eqnarray}
%The use of the formula $d\wedge (K^{-1}dK)=-K^{-1}dK\wedge K^{-1}dK$
which transform into the Cartan group structure equations after using (\ref{comr})
\begin{eqnarray}
d\wedge\omega^{k}=\frac{1}{2}c^{k}{}_{ij}\omega^{i}\wedge\omega^{j} -  c^{k}{}_{i\beta}\theta^{\beta}\wedge\omega^{i}, 
\label{Tor} \\
d\wedge\theta^{\alpha}= \frac{1}{2}c^{\alpha}{}_{ij}\omega^{i}\wedge\omega^{j}  
-  \frac{1}{2}c^{\alpha}{}_{\gamma\beta}\theta^{\beta}\wedge\theta^{\gamma}.
\label{curv}
\end{eqnarray}
To see that the new variables presented by the Cartan multiplets do not contain any abundant 
 {\it physical} DOF,  we note that Eqs. (\ref{Tor}-\ref{curv}) are equivalent to the
 Maurer-Cartan (M-C) equations defining the geometric characteristics of the 
space $G/H$ \cite{Car}, \cite{Vol1}. By solving these equations one can restore the N-G fields 
identified with the coordinates of  $G/H$ modulo its motion as a whole and coordinate reparametrizations.
The mentioned  M-C equations 
\begin{eqnarray}
 d\wedge\omega^{i} + \omega^{i}{}_{k}\wedge\omega^{k} =\frac{1}{2}c^{i}{}_{jk}\omega^{j}\wedge\omega^{k}, 
 \label{spaceT} \\
d\wedge\omega^{k}{}_{l} + \omega^{k}{}_{j}\wedge\omega^{j}{}_{l}=\frac{1}{2}
c^{k}{}_{l\beta}c^{\beta}{}_{ij}\omega^{i}\wedge\omega^{j}. 
\label{spaceR}
\end{eqnarray}
follow from  (\ref{Tor}-\ref{curv}) after using  the definition of the spin 
connection $\omega^{i}{}_{k}(d)$   
\begin{eqnarray}\label{spinc}
 \omega^{i}{}_{k}(d)=c^{i}{}_{k\beta}\theta^{\beta}(d)
\end{eqnarray}
and describe the spaces with constant torsion and curvature tensors 
\begin{eqnarray}\label{T,R}
T^{i}{}_{jk}=c^{i}{}_{jk}, \ \ \ \ \ \  R^{k}{}_{lij}=c^{k}{}_{l\beta}c^{\beta}{}_{ij}.
\end{eqnarray}
To obtain Eqs. (\ref{spaceT}-\ref{spaceR}) there  were used the Jacobi 
identities (\ref{Jab}).
e
The l.h.s. of Eqs.  (\ref{spaceT}-\ref{spaceR}) coincide with the covariant differentials (\ref{covd})
and (\ref{strh}), where $\theta^{\alpha}(d)$ is changed by the spin connection one-form $\omega^{i}{}_{k}(d)$.
This substitution creates the covariant differential $D\equiv D^{spin}$ 
in the curved space $G/H$ and presents its M-C Eqs. (\ref{spaceT}-\ref{spaceR})
in the equivalent form
\begin{eqnarray}
D\wedge\omega^{i}=\frac{1}{2}T^{i}{}_{jk}\omega^{j}\wedge\omega^{k}, \label{Tor'} 
\\
D\wedge\omega^{i}{}_{k}=\frac{1}{2}R^{i}{}_{kjl}\omega^{j}\wedge\omega^{l}  \label{curv'}.
\end{eqnarray} 
Eqs. (\ref{curv'}) show that the {\it physical} DOF represented by the two-form gauge
strength $F^{\alpha}$ (\ref{strh}) are covariantly expressed
 through the constant Riemannian tensor of $G/H$ projected on the one-forms $\omega^{i}$ 
of the vector Cartan multiplet
\begin{eqnarray}\label{F-R}
F^{i}{}_{j}\equiv c^{i}{}_{j\alpha}F^{\alpha}= R^{i}{}_{jkl}\omega^{k}\wedge\omega^{l}.
\end{eqnarray}  
For the torsionless Riemannian spaces with $T^{i}{}_{jk}=0$ their spin 
connection $\omega_{\nu}{}^{i}{}_{k}$ turns out
 to be expressed through $\omega_{i}$
%the metric $g_{\lambda\rho}(\pi)$ (\ref{mtr})
  by the relation \cite{Zbran}
\begin{eqnarray}\label{gtA}
\Gamma_{\nu\lambda}^{\rho}
=\omega_{i}^{\rho}\omega_{\nu}{}^{i}{}_{k}\omega^{k}_{\lambda}
+\partial_{\nu}\omega^{k}_{\lambda}\omega^{\rho}_{k}
\equiv\omega^{\rho}_{i}D_{\nu}^{spin}\omega^{i}_{\lambda}, 
\end{eqnarray}
where $\partial_{\mu}\equiv \frac{\partial}{\partial \pi^{\mu}}$, and $D^{spin}_{\nu}$ 
is the above defined covariant derivative $D_{\nu}$ (\ref{covder})  %(\ref{Tor'}) 
\begin{eqnarray}\label{csym}
D_{\mu}\omega_{\nu}^{i}\equiv 
D_{\mu}^{spin}\omega_{\nu}^{i}=
\partial_{\mu}\omega_{\nu}^{i}+ \omega_{\mu}{}^{i}{}_{k}\omega_{\nu}{}^{k}. 
\end{eqnarray} 
In their turn the DOF represented by $\omega^{j}$ are restricted by constraints (\ref{Tor'}) 
and the Lagrangian EOM for the N-G fields $\pi^{\lambda}(x)$ considered below.

The change-over from the covariant exterior differentials to the covariant derivatives 
  in Eqs. (\ref{Tor'}-\ref{curv'}) transforms the latter
into the system of the first-order PDEs for the
Cartan multiplets $\omega_{m}^{i}$ and $\theta_{m}^{\alpha}$ 
	\begin{eqnarray}
D_{[m}\omega_{n]}^{i}=T^{i}{}_{jk}\omega_{m}^{j}\omega_{n}^{k},
%\frac{1}{2}c^{k}{}_{ij}\omega_{[m}^{i}\omega_{n]}^{j}, 
\label{Torc} 
\\
F_{mn}{}^{i}{}_{j}= R^{i}{}_{jkl}\omega_{m}^{k}\omega_{n}^{l}, 
\label{curvc}
\end{eqnarray}
where $[.m .n]$	means antisymmetrization in the space-time indices $m,n$.  

So, constraints (\ref{Torc}-\ref{curvc}) provide the balance between 
the {\it physical} DOF represented by the Cartan multiplets and by 
the coset fields $\pi^{\lambda}$.

Now we discuss the representation of the EOM discussed in
 Section 3 in terms of the Cartan multiplets.
 As an example, we consider action (\ref{int})  %for the N-G field and 
and after  taking into account  (\ref{gtA}) find the expression of 
geodesic Eqs. (\ref{geod}) in terms of the Cartan multiplets. 
The substitution of $\Gamma_{\nu\lambda}^{\rho}$ (\ref{gtA}) 
into (\ref{geod}) yields 
	$$
	\partial^{m}\pi^{\rho}_{,m} + \omega^{\rho}_{i} \pi^{\lambda, m}(D_{\lambda}\omega^{i}_{\sigma})\pi^{\sigma}_{, m}=0,
	$$
where $\pi^{\rho}_{,m}\equiv \partial_{m}\pi^{\rho}$,  and we obtain 
	$$
	\partial^{m}\pi^{\rho}_{,m} + \omega^{\rho}_{i} \pi^{\lambda, m}D_{\lambda}\omega^{i}_{m} - 
	\pi^{\lambda ,m}\partial_{\lambda}\pi^{\rho}_{,m}=0
	$$ 
after taking into account relations (\ref{compl}), (\ref{invers'}). 
The first and the third terms in the preceeding equation are mutually cancelled in view of 
 the relation $D_{m}\omega^{i m} =\pi^{\lambda}_{, m}D_{\lambda}\omega^{i m}$, and 
EOM (\ref{geod}) transform into the gauge covariant condition for the massless 
	vector multiplet $\omega^{i}_{m}$ 
	\begin{eqnarray}\label{geodC}
	D_{m}\omega^{i  m}=0.
	\end{eqnarray} 
Eq. (\ref{geodC}) gives an example of the additional constraints implied by the EOM 
generated by the action (\ref{intg}) quadratic in $\partial_{m}\pi^{\lambda}$. 
Thus, the  N-G bosons $\pi^{\lambda}$, parametrizing the {\it symmetric} space $G/H$ 
and described by $S$ (\ref{intg}), can be equivalently described by the massless vector 
multiplet $\omega^{i}_{m}$ interacting with the Yang-Mills field $\theta^{\alpha}_{m}$ %of  $H$ 
provided the constraints (\ref{Torc}-\ref{curvc}), (\ref{geodC}) 
\begin{eqnarray}\label{constr} 
D_{[m}\omega_{n]}^{i}=0,
\nonumber \\
F_{mn}{}^{i}{}_{j}=c^{i}{}_{j\beta}c^{\beta}{}_{kl}\omega_{m}^{k}\omega_{n}^{l}, 
\\
D^{m}\omega^{i}_{m}=0.
\nonumber
\end{eqnarray}
are satisfied.
The conclusion can be extended to  more general Lagrangians
 including the higher order invariants, 
because constraints  (\ref{Torc}-\ref{curvc}) do not depend on the Lagrangian choice. 
On the contrary,  EOM depend on the Lagrangian
 and, therefore yield new constraints instead of (\ref{geodC}) together with a 
 new  gauge invariant action expressed in terms of the Cartan multiplets. 
  
The above statement accompanied with the gauge invariant action was observed
by Faddeev and Semenov-Tyan-Shanskii in \cite{FST}, where they started from 
 the {\it invariant} Cartan forms (\ref{cafo}).  However, instead of fixing the 
coordinate frame by the above-used condition $b=0$, explicitly removing 
non-physical N-G fields, they extended the global left symmetry (\ref{actn}) 
of the phenomenological Lagrangian by its {\it gauge} symmetry under 
the {\it right} multipications $G' = Gh$ with $h \in H$.
In view of this gauge invariance the redundant N-G fields associated with 
the parameters of $H$ turned out to be non-physical DOF removed by gauge fixing. 
The use of $G$-invariant Cartan forms (\ref{cafo}) as building blocks in 
effective gauge Lagrangians ensures their invariance under the global 
left $G$-multiplications. But they have also to be invariant 
under the right gauge symmetry to preserve the number of the original DOF. 

The transformation rules of $\omega_{G}^{i}, \, \theta_{G}^{\alpha}$ 
(\ref{cafo}) under the right gauge shifts 
\begin{eqnarray}\label{rgtr}
G' = G h \ \ \ \ \rightarrow \ \ \ \  G'^{-1}dG'= h^{-1}(G^{-1}dG)h + h^{-1}dh
\end{eqnarray}
have the form of the following gauge transformations 
\begin{eqnarray}\label{fadtr}
\omega_{G}^{' i}X_{i} =\omega_{G}^{i}h^{-1}X_{i}h, \ \ \ 
\theta_{G}^{' \alpha}Y_{\alpha}= \theta_{G}^{\alpha}h^{-1}Y_{\alpha}h -ih^{-1}dh.
\end{eqnarray}
The use of (\ref{comr}) makes it possible to rewrite 
relations (\ref{fadtr}) in the form 
\begin{eqnarray}\label{inffad}
\delta\omega_{G}^{k} =-c^{k}{}_{i\beta}\epsilon^{\beta}\omega_{G}^{i},  
\ \ \ \ \ \ \
\delta\theta_{G}^{\gamma}= d\epsilon^{\gamma} 
- c^{\gamma}{}_{\alpha\beta}\epsilon^{\beta}\theta_{G}^{\alpha}
\end{eqnarray}
of the infinitesimal transformations $h\approx 1 + i\epsilon^{\beta}Y_{\beta}$
with the space-time dependent parameters $\epsilon^{\beta}$ differing 
from (\ref{inftrf}) by the change of the sign 
 $\, \epsilon^{\beta} \rightarrow -\epsilon^{\beta}$. 

This proves equivalence between the standard description 
of N-G bosons [24-27], where they are realized as coordinates 
of some $G/H$, and their description as the  Cartan multiplets 
of the right gauge group $H$. Using this 
statement we will show that $p$-brane action \cite{Zbran} 
in $D$-dimensional Minkowski space is interpreted in terms of 
fields for the spontaneously broken Poincare symmetry  $ISO(D-1)$
with the coset $ISO(1,D-1)/ISO(1,p)\times SO(D-p-1)$.
 Thereat, the Cartan formalism of moving frames discussed below 
works as a key mathematical implement.

\section {Moving frame in Minkowski space}

 The moving frame in the  Minkowski 
space $\mathbf{R}^{1,D-1}$ with the global 
coordinates $\mathbf{x}=\{x^{m}\}, (m=0,1,...,D-1)$  
is formed by the orthonormal vectors  $\mathbf{n}_{A}(\mathbf{x})$   
 \begin{eqnarray}
 \mathbf{n}_{A}(\mathbf{x})\mathbf{n}_{B}(\mathbf{x})=\eta_{AB}, 
 \ \ \ (A,B=0,1,..,D-1) \label{mfra} , \\
 d\mathbf{x}= \omega^{A}(d)\mathbf{n}_{A},  \ \ \ \  
d\mathbf{n}_{A}=-\omega_{A}{}^{B}(d)\mathbf{n}_{B}  \ \ \ \ \ \  
\nonumber
 \end{eqnarray}
with their  vertex at the point $\mathbf{x}$. 
In mathematics a frame is defined as the 
pair  $(\mathbf{x}, \mathbf{n}_{A}(\mathbf{x}))$ called 
the moving $D$-hedron. The Cartan differential 
forms $\omega^{A}(d)$ and $\omega_{A}{}^{B}(d)$ emerge 
in the form of  infinitesimal {\it translational } 
and {\it rotational} componets of the $D$-hedron, respectively.    
 In view of the $\frac{1}{2}D(D+1)$ constraints (\ref{mfra}) one can
 interprete the $D^2$ dependent componets $n_{mA}(\mathbf{x})$ 
 of the vectors $\mathbf{n}_{A}$  as elements of a 
pseudoorthogonal matrix $\hat{n}=n_{mA}$ 
parametrized by  $\frac{1}{2}D(D-1)$ 
independent parameters $\pi^{\lambda}$.
 The latter are identified with the 
parameters of the $SO(1,D-1)$ group. Then the corresponding  
fields $\pi^{\lambda}(\mathbf{x})$ can be treated as 
%emerge as generic solution %$\mathbf{n}_{A}=\mathbf{n}_{A}(\pi^{\lambda})$
%of constraints  (\ref{mfra}), and can be treated 
the N-G fields  of the completely broken $SO(1,D-1)$ symmetry.
So, the frame $\mathbf{n}_{A}(\mathbf{x})$  
gives an {\it implicit} representation of the N-G 
fields of the maximally broken Lorentz symmetry when $H=I$.
 Non-linear transformations of $\pi^{\lambda}$%fields{\it implicitly given}
 under the {\it left} multiplications
from the global Lorentz group  $SO(1,D-1)$ follow 
from the matrix multiplication 
 \begin{eqnarray}\label{Glob}
n'_{mA}=l_{m}{}^{k}n_{kA},
\end{eqnarray}
where $\hat{l}$ is a Lorentz 
transformation matrix $l_{m}{}^{p}l^{n}{}_{p}=\delta_{m}{}^{n}$. 
Having found the transformation law for $\pi^{\lambda}(\mathbf{x})$
from Eqs. (\ref{Glob}) one can build an invariant Lagrangian for 
the completely broken $SO(1,D-1)$ symmetry.

On the other hand, frame treated as a matrix carries a 
{\it right} index $A$, and one can consider 
the {\it right} action of a new {\it gauge} group $SO_{R}(1,D-1)$  
\begin{eqnarray}\label{Lorg}
\mathbf{n'}_{A}=
- L_{A}{}^{B}(\pi^{\lambda})\mathbf{n}_{B}, 
\ \ \ L_{A}{}^{B}L^{C}{}_{B}=\delta_{A}^{C}
\end{eqnarray}
which  preserves the constraints (\ref{mfra}).
As a result, the matrix $\hat{n}(\pi^{\lambda})$ 
is covariant under the {\it left} global 
and  {\it right local} transformations %$L^{A}{}_{B}(\pi^{\lambda})$
\begin{eqnarray}\label{L-Rtr}
n'_{mA}=-l_{m}{}^{k}n_{kB}(L^{-1})^{B}{}_{A}
\end{eqnarray} 
forming the group $SO(1,D-1)\times SO_{R}(1,D-1)$ which preserves (\ref{mfra}). 

But, involvement of $SO_{R}(1,D-1)$ as a new gauge
 symmetry for  a Lagrangain corresponding to the 
 complete symmetry breaking will allow to 
 remove all N-G fields.
However, the situation changes when the right gauge symmetry 
is chosen to be a {\it subgroup} $H_{R}$ of $SO_{R}(1,D-1)$, 
because it allows to remove only the N-G fields corresponding 
to the generators of the {\it unbroken} symmetry $H$. 
% We shall use this observation below to interptete the brane action \cite{Zbran} in terms
%of the N-G fields for the ${\it partially}$ broken Lorentz symmetry.  
Nevertheless, the discussed example of the 
complete breaking of the global $SO(1,D-1)$ symmetry is 
instructive for illustration of the connection between the frame 
and the {\it left} invariant Cartan form $\hat{n}^{-1}d\hat{n}$ 
%of this group  
\begin{eqnarray}\label{wform}
 \omega_{A}{}^{B}(d)= (\hat{n}^{-1}d\hat{n})_{A}{}^{B}
 \equiv\mathbf{n}_{A}d\mathbf{n}^{B}.
 \end{eqnarray}
 An infinitesimal {\it right} local rotation from $SO_{R}(1,D-1)$ applied 
 to  $\mathbf{n}_{A}$ 
 \begin{eqnarray}\label{Rtr}
 \delta\mathbf{n}_{A}=\mathbf{n}_{B}\epsilon^{B}{}_ {A}  \ \ \  
 \end{eqnarray}  
results in the following gauge transformation of  $\omega_{A}{}^{B}(d)$
  \begin{eqnarray}\label{str}
  \delta\omega_{A}{}^{B}(d)=d\epsilon_{A}{}^{B} + [\omega(d), \epsilon]_{A}{}^{B}
 \end{eqnarray}
with the commutator $[\hat\omega, \hat\epsilon]$ in the  r.h.s.. 
 We see, that  the left invariant $\hat\omega$-form transforms as a  
gauge one-form under the right transformations from $SO_{R}(1, D-1)$. 
This demands the use of the covariant differential $D_{A}{}^{B}$ 
 \begin{eqnarray}\label{covde}
D_{A}{}^{B}=\delta_{A}{}^{B}d + \omega_{A}{}^{B}(d).
\end{eqnarray}
 for a vector field $\mathbf{V}= V^{A}\mathbf{n}_{A}$ transforming 
 as $\delta V_{A}= V_{B}\epsilon^{B}{}_{A}$ that results in  
 \begin{eqnarray}\label{covdif}
 \delta (DV)_{A}= (DV)_{B}\epsilon^{B}{}_{A}. 
\end{eqnarray}
Using the covariant differential (\ref{covde}) we present  (\ref{str}) 
in the standard form 
\begin{eqnarray}\label{str'}
\delta\hat{\omega}=[D,\hat{\epsilon}].
\end{eqnarray}
and the gauge covariant 2-form of the strength $F_{A}{}^{B}$
%of the strength $F_{A}{}^{B}(d)$ 
 built from $\omega_{A}{}^{B}(d)$ 
\begin{eqnarray}\label{covstr}
F_{A}{}^{B}:=D_{A}{}^{C}\wedge D_{C}{}^{B}
= (d\wedge\omega + \omega\wedge\omega)_{A}{}^{B}
\end{eqnarray}
using the exterior product of the covariant differentials.
%(\ref{covde}) 

In Secton 4 it has been explained that the description of N-G fields
as gauge fields needs to satisfy proper integrability conditions. 
In the discussed  case these conditions  
are the integrability ones for Eqs.(\ref{wform}) 
which can be rewritten as PDEs for the frame components 
\begin{eqnarray}\label{wform'}
d\mathbf{n}_{A}=-\omega_{A}{}^{B}(d)\mathbf{n}_{B}.
\end{eqnarray}
Then the integrability conditions for Eqs.(\ref{wform'}) 
are presented in the form
\begin{eqnarray}\label{integr}
d\wedge\omega_{A}{}^{B} +\omega_{A}{}^{C}\wedge\omega_{C}{}^{B}=0  
\ \ \ \rightarrow \ \ \ F_{A}{}^{B}=0.
\end{eqnarray}
Eqs. (\ref{integr}) show that the one-form $\omega_{A}{}^{B}$  treated - 
as a gauge potential for the  {\it maximal right} gauge symmetry 
 $SO_{R}(1,D-1)$ - represents the pure gauge DOF.  
 This agrees with the above made  
statement  that the addition of $SO_{R}(1,D-1)$ 
as a new Lagrange gauge symmetry for the case of the {\it complete} 
breakdown of the Lorentz group removes all rotational N-G fields.

However, in the case of $p$-brane the Lorentz symmetry 
is {\it partially} broken to its subgroup $H=SO(1,p)\times SO(D-p-1)$. 
Then the {\it right} gauge group $SO_{R}(1,D-1)$ is changed 
by its subgroup $H_{R}= SO_{R}(1,p)\times SO_{R}(D-p-1)$.  
%it is not so for the case when the Lorentz 
%symmetry is {\it partially} broken to its subgroup $H$.
Thereat, there will be excluded only the N-G fields corresponding 
to the unbroken subgroup $H$. The remaining 
 N-G bosons will be presented  by the Cartan multiplets described by 
%Just this case of {\it partial} breaking of the Lorentz symmetry 
%is realized by $p$-branes %embedded into the Minkowski space and yield 
the invariant action \cite{Zbran}.

 \section {Hypersurfaces and moving frames}

Consider a $(p+1)$-dim. world hypersurface $\Sigma_{p+1}$
embedded into the space $\mathbf{R}^{1,D-1}$ invariant under  
the global Poincare group $ISO(1,D-1)$. 
The world coordinates of $\mathbf{x}(\xi^{\mu})$ of $\Sigma_{p+1}$ 
depend on its $(p+1)$ internal coordinates $\xi^{\mu}=(\tau,\sigma^r), \, (r=1,2,..,p)$.
One can assume that $\Sigma_{p+1}$ is hyper-volume swept by a $p$-brane. 
The fields $\mathbf{n}_{A}(\xi)$  (\ref{mfra}) attached to 
$\Sigma_{p+1}$  can fix its orientation at each point $\mathbf{x}(\xi^{\mu})$.
Any fixation of the origin and orientation of  $\mathbf{n}_{A}$ means the choice 
of a vacuum state of a $p$-brane.  Using arbitrariness in orientation of 
$\mathbf{n}_{A}$ one can fix the manifold of degenerated vacuum states
 by the condition for the first $p+1$ vectors $\mathbf{n}_{i}, \ (i,k=0,1,...,p)$
 to be tangent to $\Sigma_{p+1}$ at each point $\mathbf{x}(\xi^{\mu})$. 
The residual rotational symmetry of so chosen vacuum manifold is described by 
the $SO(1,p)\times SO(D-p-1)$ of the local right rotations of $\mathbf{n}_{A}(\xi)$.  
 Requirement of this symmetry is equivalent to spontaneous breaking of
 the left global symmetry $SO(1,D-1)$ to $SO(1,p)\times SO(D-p-1)$ 
in correspondence with our previous discussion. 
Then the rotational N-G fields can be described by  
the above-discussed Cartan multiplets. 
A fixing of the order parameter
$\mathbf{x}(\xi)$ of $\Sigma_{p+1}$ will spontaneously break the global
 translational symmetry of $\mathbf{R}^{1,D-1}$. 
The subgroup $SO(1,p)$ consists of the Lorentz rotations
acting in the planes tangent to $\Sigma_{p+1}$ which are 
added by independent $ SO(D-p-1)$ rotations acting in 
 $(D-p-1)$-dim. subspaces normal to $\Sigma_{p+1}$. The subspaces 
are spanned by the orts $\mathbf{n}_{a} \ (a,b=p+1,p+2,..., D-p-1)$.
Thus, our choice of the vacuum manifold splits the frame 
vectors $\mathbf{n}_{A}$ (\ref{mfra}), originally encoding 
the  $\frac{1}{2}D(D-1)$ 
N-G bosons of the broken Lorentz group, into two 
subsets: $\mathbf{n}_{A}=(\mathbf{n}_{i}, \mathbf{n}_{a})$.  
These subsets form linear representations of the right local subgroups  
$SO(1,p)$ and $SO(D-p-1)$, respectively. As a result, among the $\frac{1}{2}D(D-1)$ 
rotational N-G bosons there are  $\frac{1}{2}(p+1)p +\frac{1}{2}(D-p-1)(D-p-2)$  
representing non-physical DOF which can be removed by gauge fixing. 

The remaining $(p+1)(D-p-1)$ DOF correspond to the real rotational N-G bosons  
identified with the coordinates of the coset $SO(1,D-1)/SO(1,p)\times SO(D-p-1)$,  
and can be  described in terms of the constrained Cartan vector and gauge multiplets. 
These multiplets form  the blocks of the differential matrix 
one-form $\omega_{A}{}^{B}$ (\ref{wform}) belonging to the Lorentz algebra $SO(1,D-1)$ 
\begin{eqnarray}\label{spl}
\omega_{A}{}^{B}(d)
=\left( \begin{array}{cc}
                       A_{i}{}^{k}& W_{ i}{}^{b} \\
                         W_{ a}{}^{ k} & B_{a}{}^{b}
                              \end{array} \right) .
\end{eqnarray}
The diagonal one-form submatrices  $A_{\mu i}{}^{k}d\xi^{\mu}$ and  $B_{\mu a}{}^{b}d\xi^{\mu}$ 
in (\ref{spl}) describe the gauge fields in the fundamental representations of the unbroken $SO(1,p)$
and $SO(D-p-1)$ subgroups, respectively. 
The off-diagonal submatrix $W_{\mu i}{}^{b}d\xi^{\mu}$, having $(p+1)(D-p-1)$ components equal
 to the number of N-G bosons, describes a charged vector multiplet in the bi-fundamental 
 representation of the subgroup $SO(1,p)\times SO(D-p-1)$ with the covariant derivative 
\begin{eqnarray}\label{cd}
(D_{\mu} W_{\nu})_{i}{}^{a}= \partial_{\mu}W_{\nu i}{}^{a}+ A_{\mu i}{}^{k} W_{\nu k}{}^{a} + 
B_{\mu}{}^{a}{}_{b} W_{\nu i}{}^{b}. 
\end{eqnarray}
The covariant derivatives  $D_{\mu}^{||}$ and $D_{\mu}^{\perp}$ associated with the gauge 
fields  $\hat A_{\mu}$ and $\hat B_{\mu}$, respectively,  are given by 
\begin{eqnarray}
D_{\mu}^{||}\Phi^{i}
=\partial_{\mu}\Phi^{i}+ A_{\mu}{}^{i}{}_{k}\Phi^{k}, \\
\label{||} 
D_{\mu}^{\perp}\Psi^{a}
=\partial_{\mu}\Psi^{a}+ B_{\mu}{}^{a}{}_{b}\Psi^{b} \label{perp}. 
\end{eqnarray}
In terms of these covariant derivatives the strengths $F_{\mu\nu i}{}^{k}$, $H_{\mu\nu a}{}^{b}$ 
 for ${\hat A}_{\mu}$ and  ${\hat B}_{\mu}$, respectively, are presented  as follows
\begin{eqnarray}
F_{\mu\nu i}{}^{k}\equiv 
[D_{\mu}^{||},\,  D_{\nu}^{||}]_{i}{}^{k}=(\partial_{[\mu}A_{\nu]} 
+ A_{[\mu}A_{\nu]})_{i}{}^{k} 
 \label{F} \\
H_{\mu\nu a}{}^{b}\equiv 
[D_{\mu}^{\perp},\,  D_{\nu}^{\perp}]_{a}{}^{b}=
(\partial_{[\mu}B_{\nu]} + B_{[\mu}B_{\nu]})_{a}{}^{b}. 
 \label{H}
\end{eqnarray} 
Then the  integrability conditions (\ref{integr}) 
take the form of the constraints 
\begin{eqnarray}
F_{\mu\nu i}{}^{k}= -(W_{[\mu} W_{\nu]})_{i}{}^{k},
\label{cF} \\
H_{\mu\nu a}{}^{b}= -(W_{[\mu} W_{\nu]})_{a}{}^{b},
\label{cH} \\
(D_{[\mu} W_{\nu]})_{i}{}^{a}=0 \label{ccd}
\end{eqnarray}
which are the Gauss-Ricci-Codazzi eqs. reformulated in 
terms of the massless vector multiplet $W_{\mu  i}{}^{a}$ 
and the gauge strenghts $\hat{F}_{\mu\nu}$, $\hat{H}_{\mu\nu}$ 
of the unbroken subgroup of the gauge Lorentz group
attached to the base hypersurface \cite{Zbran}.

The geometry of $\Sigma_{p+1}$
is described  by the induced pseudo-Riemannian 
metric $g_{\mu\nu}(\xi)$ defining the Levi-Chivita 
connection $\Gamma_{\mu\nu}^{\rho}$
\begin{eqnarray}\label{gmetr} 
g_{\mu\nu}(\xi)=\partial_{\mu}\mathbf{x}\partial_{\nu}\mathbf{x},\ \ \
\Gamma_{\mu\nu}^{\rho}
=\frac{1}{2}g^{\rho\gamma}(\partial_{\mu}g_{\nu\gamma} 
+ \partial_{\nu}g_{\mu\gamma}
- \partial_{\gamma}g_{\mu\nu})
\end{eqnarray}
which must be added into the gauge-covariant 
 derivative $(D_{\mu} W_{\nu})_{i}{}^{a}$  (\ref{cd}) %defined by relation
\begin{eqnarray}\label{grcd}
D_{\mu} W_{\nu i}{}^{a} \ \  \rightarrow \ \
\hat{\nabla}_{\mu}W_{\nu i}^{a}= \partial_{\mu}W_{\nu i}{}^{a}+ A_{\mu i}{}^{k} W_{\nu k}{}^{a} + 
B_{\mu}{}^{a}{}_{b} W_{\nu i}{}^{b} - \Gamma_{\mu\nu}^{\rho}W_{\rho i}^{a}
\end{eqnarray}
that makes it covariant under reparametrizations of $\Sigma_{p+1}$. 
In view of the symmetry $\Gamma_{\mu\nu}^{\rho}=\Gamma_{\nu\mu}^{\rho}$ 
the covariantization changes not the G-R-C constraints (\ref{cF}-\ref{ccd}), 
but the commutator of the derivatives (\ref{cd})
\begin{eqnarray}\label{BI}
[\hat{\nabla}_{\mu} , \, \hat{\nabla}_{\nu}] 
= \hat{F}_{\mu\nu} + \hat{H}_{\mu\nu} +\hat{R}_{\mu\nu} 
\end{eqnarray}
by adding the Riemann-Cristoffel tensor of the  metric $g_{\mu\nu}$ (\ref{gmetr})
\begin{eqnarray}\label{Riem}
R_{\mu\nu}{}^{\gamma}{}_{\lambda}
= \partial_{[\mu}\Gamma_{\nu]\lambda}^{\gamma} 
+\Gamma_{[\mu|\rho}^{\gamma}\Gamma_{|\nu] \lambda}^{\rho}.
\end{eqnarray}

The invariant Cartan forms for the generators of 
{\it global} translations of $\mathbf{R}^{1,D-1}$ are equal to $dx^{m}$, 
where $x^{m}$ are the global Cartesian coordinates 
in  Minkowski space. 
The projections of $d\mathbf{x}$ on the local frame vectors 
$\mathbf{n}_{A}(\xi)$ coincide with components of the local 
 translational Cartan form $\omega^{A}$ (\ref{mfra})
\begin{eqnarray}\label{pulb} 
\omega^{A}(d)=d\mathbf{x}\mathbf{n}^{A}(\xi)\equiv dx^{m}n_{m}{}^{A}
\end{eqnarray}
 which is the $D$-bein one-form of the
Minkowski space referred to the local frame.
 Eqs.(\ref{pulb}) show the role of $n^{A}_{m}(x)
=\frac{\omega^{A}(d)}{dx^{m}}$ as a local Lorentz 
matrix $n_{m}{}^{A}n_{nA}=\eta_{mn}$ connecting the components of the 
global and local vielbein one-forms. 
So, we can treat  (\ref{pulb}) as the PDEs which express the local
 translations $\omega^{A}(d)$ through $d\mathbf{x}$.
The integrability conditions for Eqs. (\ref{pulb})  
\begin{eqnarray}\label{intrl}
d\wedge\omega_{A}+ \omega_{A}{}^{B}\wedge \omega_{B}=0 
\end{eqnarray}
reproduce the second  set of the M-C equations for the Minkowski space  
additional to the first set (\ref{integr}). 
Eqs. (\ref{integr}) and (\ref{intrl}) form the M-C structure equations of the Minkowski space
which is the homogeneous space $ISO(1, D-1)/SO(1, D-1)$  for the global Poincare symmetry.  

After localization of the  translation parameters $\epsilon^{m}(\xi^{\mu})$   
on  $\Sigma_{p+1}$  the transformation law for  $dx^{m}(\xi)$  changes into the form 
\begin{eqnarray}\label{trnsl} 
\delta dx^{m} =d\epsilon^{m}
\end{eqnarray}
like an abelian gauge one-form.
The splitting $\mathbf{n}_{A}=(\mathbf{n}_{i}, \mathbf{n}_{a})$ induces the 
splitting $\omega_{A}(\xi)=(\omega_{i}, \omega_{a})$ into the tangent and orthogonal
 components to $\Sigma_{p+1}$ at each of its points.  
 In the chosen vacuum manifold $\mathbf{n}_{i}$ are tangent to $\Sigma_{p+1}$ 
that is equivalent to the orthogonality conditions  
\begin{eqnarray}\label{trl} 
\omega_{a}\equiv\mathbf{n}_{a}(\xi)d\mathbf{x}(\xi)=0 \ \ \ \rightarrow \ \
d\mathbf{x}=\omega^{i}(d)\mathbf{n}_{i}(\xi). 
\end{eqnarray}
In view of invariance of  $\omega_{a}(d)$ under  the left  $ISO(1,D-1)$ global
transformations one can, of course, interpret the conditions $\omega_{a}=0$ 
similarly to the inverse Higgs phenomenon \cite{IO}.
 In our case, however, these conditions mean the choice 
 of $ISO(1, p)\times SO(D-p-1)$ as a
 vacuum subgroup for the Poincare symmetry 
 expressing physical equivalence of the remaining 
orientations and positions of $\Sigma_{p+1}$  in the Minkowski space. 
Invariance of (\ref{trl})  under 
the {\it right local} $ISO(1, p)\times SO( D-p-1)$ transformations 
 of the moving polyhedron $(\mathbf{x}(\xi), \mathbf{n}_{A}(\xi))$ 
 shows that the Poincare symmetry is spontaneously broken.

In view of  $\omega_{a}=0$ the quadratic element 
$ds^{2}=d\mathbf{x}^2$ of $\Sigma_{p+1}$ takes the form
\begin{eqnarray}\label{metr}
ds^{2}=\omega_{i}\omega^{i}
=\omega^{i}_{\mu}\omega_{i\nu}d\xi^{\mu}d\xi^{\nu}\equiv g_{\mu\nu}(\xi)d\xi^{\mu}d\xi^{\nu}. 
\end{eqnarray}
This shows that the values $\omega^{i}_{\lambda}(\xi)$ are the components of a $(p+1)$-bein 
 of $\Sigma_{p+1}$ in terms of which its induced metric $g_{\mu\nu}(\xi)$ is 
\begin{eqnarray}\label{metric}
g_{\mu\nu}:=\mathbf{e}_{\mu}\mathbf{e}_{\nu}
=\omega_{\mu}^{i}\eta_{ik}\omega_{\nu}^{k},
\end{eqnarray} 
where $\mathbf{e}_{\mu}(\xi)$ is the natural frame on  $\Sigma_{p+1}$
\begin{eqnarray}\label{natf}
\mathbf{e}_{\mu}=\omega_{\mu}^{i}\mathbf{n}_{i}, \ \ \ \omega_{\mu}^{i}\omega^{\mu}_{k}=\delta^{i}_{k}.
\end{eqnarray} 
 So, the unbroken translational modes from  $\mathbf{x}(\xi)$ are condensed in
 the $(p+1$)-bein $\omega^{i}_{\lambda}$ and the metric $g_{\mu\nu}$.
The dynamical role of the broken translational modes becomes clear after the
  substitution of $(D-p-1)$ conditions $\omega^{a}=0$ into the integrability 
  conditions (\ref{intrl}) that results in their splitting 
\begin{eqnarray}
D_{[\mu}^{||}\omega_{\nu]}^{i}\equiv
\partial_{[\mu}\omega_{\nu]}^{i}+ A_{[\mu}{}^{i}{}_{k}\omega_{\nu]}{}^{k}=0, 
 \label{csym'} \\
\omega_{[\mu}^{i}W_{\nu]ia}=0 \label{lcntr} \ \ \ \ \ \ \ \ \ \ \ \ 
\end{eqnarray}
connecting $\omega_{\mu}^{i}$ with the fields  $W_{\mu i}{}^{a}$ 
and $A_{\mu}{}^{i}{}_{k}$  (\ref{cd}).

The solution of constraints (\ref{lcntr}) 
\begin{eqnarray}\label{2frm}
W_{\mu i}{}^{a}= -l_{\mu\nu}{}^{a}\omega^{\nu}_{i}, \ \ \ \
\end{eqnarray}
reveals that the rotational N-G modes $\hat{W}_{\mu}$ 
referred to the natural frame $\mathbf{e}_{\mu}(\xi)$ (\ref{natf})
turn out to be components of the second fundamental form 
 $l_{\mu\nu}{}^{a}$  of $\Sigma_{p+1}$ 
\begin{eqnarray}\label{secfor}
l_{\mu\nu}{}^{a}:=\mathbf{n}^{a}\partial_{\mu\nu}\mathbf{x}.    
\end{eqnarray}
Thus, we find that the fixation of the vacuum manifold by the
 conditions $\omega^{a}=0$ makes the N-G bosons of the broken 
 translations shifted into the second fundamental form. 
  This permits to express the N-G bosons of the broken 
Lorentz transformations and translations through the components 
of $l^{a}_{\mu\nu}$. 

Another effect of the vacuum conditions follows from Eqs. (\ref{csym'}) which
are equivalent to the conditions of the parallel transport of  $\omega^{i}_{\mu}$ 
along  $\Sigma_{p+1}$
\begin{eqnarray}\label{tetpo}
\nabla_{\mu}^{||}\omega_{\nu}^{i}
\equiv\partial_{\mu}\omega_{\nu}^{i} - \Gamma_{\mu\nu}^{\rho}\omega_{\rho}^{i}
 + A_{\mu}{}^{i}{}_{k}\omega_{\nu}^{k} = 0 
\end{eqnarray}
which manifests the tetrade postulate resulting in solution (\ref{gtA}). 
The latter represents  $\Gamma_{\mu\nu}^{\rho}$ through the gauge 
field $A_{\mu}{}^{i}{}_{k}$. The inverse relation represents   $A_{\mu}{}^{i}{}_{k}$
and its strength  $F_{\mu\nu i}{}^{k}$ through 
  $\Gamma_{\nu\lambda}^{\rho}$ and the Riemann  
  tensor $R_{\mu\nu}{}^{\gamma}{}_{\lambda}$ (\ref{Riem}) 
 \begin{eqnarray}\label{gtA'}
  A_{\nu}^{lm}= \omega^{l}_{\rho}\Gamma_{\nu\lambda}^{\rho}\omega^{\lambda m} +
\omega^{l}_{\lambda}\partial_{\nu}\omega^{\lambda m}, \\
F_{\mu\nu}{}^{i}{}_{k}= 
\omega^{i}_{\gamma} R_{\mu\nu}{}^{\gamma}{}_{\lambda}\omega^{\lambda}_{k}.\ \ \ \ \ \
\label{FRcon}
\end{eqnarray}
This means that the $SO(1,p)$ gauge field referred to the natural 
frame $\mathbf{e}_{\mu}(\xi)$ turns out to  
be  the metric connection $\Gamma_{\nu\lambda}^{\rho}$. 
The substitutions of $\Gamma_{\nu\lambda}^{\rho}$ instead of 
the gauge field  $A_{\nu ik}$, and the tensor field  
$l_{\mu\nu}{}^{a}= -\omega_{\nu}^{i}W_{\mu i}{}^{a}$ for 
  $W_{\mu i}{}^{a}$ into the  G-R-C constraints (\ref{cF}-\ref{ccd}) transform them into 
	the desired constraints \cite{Zbran}
\begin{eqnarray}
R_{\mu\nu}{}^{\gamma}{}_{\lambda}=l_{[\mu}{}^{\gamma a} l_{\nu]\lambda a},
\label{cRl} \\
H_{\mu\nu }{}^{ab}= l_{[\mu}{}^{\gamma a} l_{\nu]\gamma}{}^{b},
\label{cH2} \\ 
\nabla_{[\mu}^{\perp}l_{\nu]\rho a}=0, \ \ \ \ \ \
\label{ccd'}
\end{eqnarray}
where the general and  $SO(D-p-1)$ covariant  
derivative  $\nabla_{\mu}^{\perp}$ is defined as 
 \begin{eqnarray}\label{cdl}
\nabla_{\mu}^{\perp}l_{\nu\rho}{}^{a}:= \partial_{\mu}l_{\nu\rho}{}^{a}
- \Gamma_{\mu\nu}^{\lambda} l_{\lambda\rho}{}^{a} 
-\Gamma_{\mu\rho}^{\lambda} l_{\nu\lambda}{}^{a} + B_{\mu}^{ab}l_{\nu\rho b}.
 \end{eqnarray} 
The commutator of these covariant derivatives %$\nabla^{\perp}{}_{\mu}$ (\ref{cdl})
 yields the Bianchi identitities
  \begin{eqnarray}\label{BIl}
[\nabla^{\perp}_{\gamma} , \, \nabla^{\perp}_{\nu}] l^{\mu\rho a}
=R_{\gamma\nu}{}^{\mu}{}_{\lambda} l^{\lambda\rho a}  
+ R_{\gamma\nu}{}^{\rho}{}_{\lambda} l^{\mu\lambda a} 
+H_{\gamma\nu}{}^{a}{}_{b} l^{\mu\rho b}.  
\end{eqnarray}

 Eqs. (\ref{cRl}) generalize the $\it {Gauss \, Theorema  \, Egregium}$ 
for surfaces in 3-dim. Euclidean space  to the case 
of (p+1)-dimensional  world hypersurfaces  
embedded into $D$-dim. Minkowski space. 

To identify $\Sigma_{p+1}$ with the $p$-brane hypervolume we have 
 to construct a gauge invariant  $(p+1)$-dim. action formulated in terms of 
the Cartan multiplets and to prove that its Euler-Lagrange EOM are compatible with 
constraints (\ref{cRl}-\ref{ccd'})  and the EOM of a $p$-brane. 
The problem is solved in the next section according to 
the scheme described in Section 4.

\section {Gauge invariant action for branes}

The  Dirac $p$-brane action \cite{Dir}
in $D$-dimensional Minkowski space equals the brane hypervolume  
including the determinant of the induced metric $g_{\mu \nu}$ (\ref{gmetr}) 
\begin{equation}\label{Dirbr}
S=T\int d^{p+1}\xi \sqrt{|det(\partial_{\mu}\mathbf{x}\partial_{\nu}\mathbf{x})|}.
\end{equation}
This action yields the non-linear wave equation for the world vector $\mathbf{x}(\xi^{\mu})$ 
\begin{equation}\label{Box}
\Box^{(p+1)}\mathbf{x}=0, 
\end{equation}
where $\Box^{(p+1)}$ is the D'Alembert-Beltrami operator on the hypersurface $\Sigma_{p+1}$ 
\begin{equation}\label{LaBe}
\Box^{(p+1)}:
=\frac{1}{\sqrt{|g|}}\partial_{\alpha} \sqrt{|g|}g^{\alpha\beta}\partial_{\beta}.
\end{equation} 
 Eq. (\ref{Box}) means that the Dirac $p$-brane sweeps the minimal 
 hypersurface $\Sigma_{p+1}$ defined  by the algebraic conditions 
\begin{equation}\label{minco}
Spl^{a}\equiv g^{\mu\nu}l^{a}_{\mu\nu}=0
\end{equation} 
for the second fundamental form $l_{\mu\nu}{}^{a}$ defined 
by (\ref{secfor}). The equivalence of these conditions 
to the wave Eq. (\ref{Box}) follows from  Eqs. (\ref{trl}) as 
a result of the orthogonality between  $\mathbf{n}^{a}$ and the 
 vectors $\partial_{\beta}\mathbf{x}$ tangent to $\Sigma_{p+1}$:
  $\mathbf{n}^{a}\frac{\partial\mathbf{x}}{\partial\xi^{\beta}}=0$ 
that makes the metric connection contribution vanishing.
  
 In correspondence with the gauge theory reformulation of the 
broken Lagrangians discussed in Section 4 
 the minimality condition (\ref{minco}) plays the role similar to 
Eq. (\ref{geodC}) representing the standard  N-G  EOM  in the gauge approach. 
 So,  Eqs. (\ref{cRl}-\ref{ccd'}), (\ref{minco}) are analogous 
 to Eqs.  (\ref{constr}), 
 and yield a complete set of data to represent 
 Dirac action (\ref{Dirbr}) in terms of the Cartan multiplets. 
 Such a representation is given by the below-discussed  gauge-invariant 
 action $S_{Dir}$  (\ref{actnl}) 
 including a brane metric $g_{\mu\nu}$
 which is considered as a given (background) field \cite{Zbran}. However, 
 it does not mean that $g_{\mu\nu}$ is not a dynamical field, because its 
 dynamics is already fixed by the Gauss condition (\ref{cRl}) treated 
 as the second order PDEs for $g_{\mu\nu}$ with a 
 given $l_{\mu\nu}^{a}$. In its turn the dynamics of the N-G 
 multiplet $l_{\mu\nu}^{a}$ is derived using the standard 
 variational principle for the action (\ref{actnl}).
For the Nambu-Goto string (p=1) in 3-dimensional Minkowski
 space the metric  condition (\ref{cRl}) is the Lioville equation on  
 string world-sheet showing that it is 2-dim. Einstein space
   $$ R_{\mu\nu}=\frac{R}{2}g_{\mu\nu}.$$
 Keeping in mind such peculiarities in application 
 of the variational principle for dynamical description 
 of the metric and  Cartan fields we consider derivation of (\ref{actnl}) 
 step by step.  It will permit to avoid a possible missunderstanding of the 
 presented results. 
 
 In the long wave approximation the leading order for a gauge invariant brane
 action has to be quadratic in the derivatives of the Cartan 
 multiplets. To find such an action we start from a generalized 
 Landau-Ginzburg action
\begin{eqnarray}
S= \gamma\int d^{p+1}\xi\sqrt{|g|}  
\{ - \frac{1}{4}Sp(H_{\mu\nu}H^{\nu\mu})\nonumber 
+ \frac{1}{2}\nabla_{\mu}^{\perp}l_{\nu\rho a}\nabla^{\perp \{\mu}l^{\nu\}\rho a}
\\
-\nabla_{\mu}^{\perp}l^{\mu}_{\rho a}\nabla^{\perp}_{\nu}l^{\nu\rho a}  + V \} 
\ \ \ \ \ \ \ \ \ \ \ \ \ \ \ \
\label{actn2}
\end{eqnarray}
with a potential $V$ encoding self-interaction of the N-G fields $l_{\mu\nu}^{a}$ 
and compatible with  Eqs. (\ref{cRl}-\ref{ccd'}).
We prove that Eqs. (\ref{cRl}-\ref{ccd'}), (\ref{minco}) permit to restore 
 $V$ in a unique fashion.
The EOM for $B_{\mu}{}^{ab}$ and $l_{\mu\nu}{}^{a}$ 
following from  (\ref{actn2}) are
\begin{eqnarray}
\nabla^{\perp}_{\nu} {\cal H}^{\nu\mu}_{ab}= 
 \frac{1}{2}l_{\nu\rho[a}\nabla^{\perp[\mu} l^{\nu]\rho}{}_{b ]},
  \ \ \ \ \ \ \ \ \ \ \ \ \ \
\label{maxH2'} \\
\frac{1}{2}\nabla^{\perp}_{\mu}\nabla^{\perp[\mu}l^{\{\nu]\rho\}a}
=-[\nabla^{\perp\mu},  \nabla^{\perp\{\nu}] l_{\mu}{}^{\rho\}a}
+ \frac{\partial {V}}{\partial l_{\nu\rho a}},   
\label{eqgW2ss}
\end{eqnarray}
where  ${\cal H}_{\mu\nu}^{ab}= H_{\mu\nu}^{ab} 
- l_{[\mu}{}^{\gamma a} l_{\nu]\gamma}{}^{b}$ 
is the shifted strenght $H_{\mu\nu}^{ab}$. 

Then one can see that the Ricci and Codazzi 
eqs.  (\ref{cH2}-\ref{ccd'}) presented as
\begin{eqnarray}\label{Hlsol} 
{\cal H}_{\mu\nu}^{ab}=0, \ \ \ \ \  
\nabla_{[\mu}^{\perp}l_{\nu]\rho}^{a}=0, 
\end{eqnarray} 
 form the solution of Eqs. (\ref{maxH2'}-\ref{eqgW2ss}) provided 
  $V$ satisfies the conditions 
\begin{eqnarray}\label{eqV} 
 \frac{\partial {V}}{\partial l_{\nu\rho a}}
=[\nabla^{\perp\mu},  \nabla^{\perp\{\nu}] l_{\mu}{}^{\rho\} a}.       
\end{eqnarray}
The Bianchi identitities (\ref{BIl}) permit to express the r.h.s. of (\ref{eqV})
through $R_{\gamma\nu\rho\lambda}$,  
$H_{\gamma\nu}^{ab}$ and $l_{\mu\nu}^{a}$.
Then conditions ({\ref{eqV}) could be transformed into PDEs 
 for the scalar function $V(l)$ if  $R_{\gamma\nu}{}^{\rho}{}_{\lambda}$ 
 and $H_{\gamma\nu}{}^{ab}$ could be represented  as some  
explicit functions of  $l_{\mu\nu}^{a}$. 
Such a representation is given by the Gauss (\ref{cRl}) 
and Ricci  (\ref{cH2}) conditions. This clarifies 
 the dynamical role of these conditions as the selection 
rules {\it defining} $V$ and the {\it metric} in $S$  (\ref{actn2}). 
The substitution of Eqs. (\ref{cRl}-\ref{cH2}) in (\ref{eqV}) transforms the
latter into the PDEs for the potential $V(l)$ {\it fixing} it to be the 
invariant {\it quartic} polynomial in $l_{\mu\nu}{}^{a}$ 
\begin{eqnarray}\label{eqVl}
\frac{1}{2} \frac{\partial {V}}{\partial l_{\nu\rho a}}
=\frac{1}{2}(l^{\{a}l^{b\}})^{\rho\nu}Spl_{b} 
+ (2l_{b}l^{a}l^{b}-l^{a}l_{b}l^{b}-l_{b}l^{b}l^{a})^{\rho\nu} 
-l^{\rho\nu}_{b}Sp(l^{b}l^{a}) 
\end{eqnarray}
where $Sp(l^{a}l^{b})=g^{\mu\nu}l^{a}_{\mu\rho}l^{b\rho}_{\nu}$. 
The solution corresponding to Dirac branes is 
\begin{eqnarray}\label{solVl} 
V_{Dir}=- \frac{1}{2} Sp(l_{a}l_{b}) Sp(l^{a}l^{b})
+ Sp(l_{a}l_{b}l^{a}l^{b}) - Sp(l_{a}l^{a}l_{b}l^{b})+c,  \ \  Spl^{a}=0
\end{eqnarray}
with the integration constant $c$. 
The minimality condition in (\ref{solVl}) is invariant under all 
symmetries of S (\ref{actn2}), and is interpreted 
as the inverse Higgs phenomenon condition additional to $\omega_{a}=0$. 
 Then EOM (\ref{eqgW2ss}) reduce to
\begin{eqnarray}\label{eqgW2sr} 
\frac{1}{2}\nabla^{\perp}_{\mu}\nabla^{\perp[\mu}l^{\{\nu]\rho\}a}\equiv
\nabla^{\perp}_{\mu}(\nabla^{\perp[\mu}l^{\nu]\rho a} 
 + \frac{1}{2}\nabla^{\perp[\nu}l^{\rho]\mu a})=0  \rightarrow \  
 \nabla^{\perp}_{\rho}\nabla^{\perp}_{\mu}l^{\mu\rho a}=0
 \end{eqnarray}
 where we used the identity $\nabla^{\perp[\mu}l^{[\nu]\rho]}_{a}
=-\nabla^{\perp[\nu}l^{\rho]\mu}_{a}$.

As a result, we obtain  the desired invariant action for 
 the Dirac $p$-brane 
 \begin{eqnarray}
S_{Dir}= \gamma\int d^{p+1}\xi\sqrt{|g|} \, \, \mathcal{L} ,  \nonumber 
  \ \  \ \  \ \ \ \  \ \  \ \  \ \ \ \  \ \  \ \  \ \ \ \  \ \  \ \ 
\\
\mathcal{L}= - \frac{1}{4}Sp(H_{\mu\nu}H^{\nu\mu}) 
+ \frac{1}{2}\nabla_{\mu}^{\perp}l_{\nu\rho a}\nabla^{\perp \{\mu}l^{\nu\}\rho a}
-\nabla_{\mu}^{\perp}l^{\mu}_{\rho a}\nabla_{\nu}^{\perp}l^{\nu\rho a} 
\nonumber 
\\
- \frac{1}{2} Sp(l_{a}l_{b}) Sp(l^{a}l^{b})
+ Sp(l_{a}l_{b}l^{a}l^{b})
- Sp(l_{a}l^{a}l_{b}l^{b})   + c  \ \  \ \  \ \  \ \  \ \  \ \  \  \ \  \ \ 
\label{actnl} 
\end{eqnarray}
desribing the interacting traceless tensor  $l_{\mu\nu}^{a}$ and 
vector $B_{\mu}^{ab}$ multiplets in the background metric $g_{\mu\nu}$.  
Thereat, the dynamical equations for  $g_{\mu\nu}$ are encoded by the 
Gauss conditions (\ref{cRl}) automatically built in $S_{Dir}$.
 
 The Euler-Lagrange EOM for $S_{Dir}$ (\ref{actnl}) 
 can be written in the form
 \begin{eqnarray}
\nabla^{\perp}_{\nu} {\cal H}^{\nu\mu}_{ab}
= \frac{1}{2}l_{\nu\rho[a}\nabla^{\perp[\mu} l^{\nu]\rho}{}_{b ]},
  \ \ \ \ \ \ \ \ \ \ \ \ \ \
\label{maxH2*} \\
\nabla^{\perp}_{\mu}\nabla^{\perp[\mu}l^{\nu]\rho a} + 
\frac{1}{2}\nabla^{\perp[\nu}\nabla^{\perp}_{\mu}l^{\rho]\mu a} =0,  
\ \ \ \ \ \ \ \ \  
\label{eqgW2*}
\end{eqnarray}
of the {\it second order} PDEs after using Eqs. (\ref{minco}), (\ref{eqgW2sr})
 and the identities   
\begin{eqnarray}\label{BIls}
[\nabla^{\perp\mu} , \, \nabla^{\perp[\nu}]l_{\mu}^{\rho]a}
=([l^{b},l^{a}])^{\nu\rho} Spl_{b}
\end{eqnarray}
which follow from the Bianchi identities (\ref{BIl}).
 
 Now we prove that Eqs. (\ref{Hlsol}) yield the {\it general solution} of 
 (\ref{maxH2*}-\ref{eqgW2*}).
 Taking into account that (\ref{Hlsol}) 
are PDEs of the {\it first order} we consider them to be the {\it Cauchy initial 
data} for PDEs (\ref{maxH2*}-\ref{eqgW2*}) chosen at the time  $\tau=0$
 \begin{eqnarray}\label{cauchy}
{\cal H}^{\nu\mu}_{ab}(0,\sigma^r)=0, \ \ \ \
 \nabla^{\perp[\mu}l^{\nu]\rho}_{a}(0,\sigma^r)=0,   
 \ \ \ \ \nabla^{\perp}_{\mu}l^{\mu\rho}_{a}(0,\sigma^r)=0 
\end{eqnarray} 
and show that the R-C Eqs. (\ref{Hlsol}) are always satisfied in view of 
 EOM (\ref{maxH2*}-\ref{eqgW2*}). Using the power series expansion of 
 ${\cal H}^{\nu\mu}_{ab}$ and $\nabla^{\perp[\mu}l^{\nu]\rho a}$
 \begin{eqnarray}
 {\cal H}^{\tau r}_{ab}(\delta\tau,\sigma^r)={\cal H}^{\tau r}_{ab}|_{\tau=0}%(0,\sigma^r ) 
 + \partial_{\tau} {\cal H}^{\tau r}_{ab}|_{\tau=0}\delta\tau + ...
 =\nabla^{\perp}_{\tau}{\cal H}^{\tau r}_{ab}|_{\tau=0}\delta\tau + ... ,  \label{varid}  
 \\
 \nabla^{\perp[\tau}l^{\nu]\rho a}(\delta\tau,\sigma^r )=
 %\nabla^{\perp[\tau}l^{\nu]\rho a}|_{\tau=0}%%(0,\sigma^r )
 \partial_{\tau} \nabla^{\perp[\tau}l^{\nu]\rho a}|_{\tau=0}\delta\tau +...
 =\nabla^{\perp}_{\tau}\nabla^{\perp[\tau}l^{\nu]\rho a}|_{\tau=0}\delta\tau +... \, .
 \nonumber 
 \end{eqnarray}
and Eqs. (\ref{maxH2*}-\ref{eqgW2*}), (\ref{cauchy}) we obtain 
 \begin{eqnarray}
 {\cal H}^{\tau r}_{ab}(\delta\tau,\sigma^r)= %{\cal H}^{\tau r}_{ab}(0,\sigma^r ) \partial_{\tau} 
 -\nabla^{\perp}_{r'} {\cal H}^{r' r}_{ab}|_{\tau=0}\delta\tau +... \, ,  \ \ \ \ \ \ \ \ \ \ \ \ \ \ \ \ \
 \label{varid'} \\
 \nabla^{\perp[\tau}l^{\nu]\rho a}(\delta\tau,\sigma^r)
 =-(\nabla^{\perp}_{r'}\nabla^{\perp[r'}l^{\nu]\rho a}
 +  \frac{1}{2}\nabla^{\perp[\nu}\nabla^{\perp}_{\mu}l^{\rho]\mu a})|_{\tau=0} \delta\tau 
  +... \, .\, .
  \nonumber 
 \end{eqnarray}
  Then observing that the space covariant derivatives %( $\nabla^{\perp}_{r'} with r'=1,2,...,p-1$ ) 
 of (\ref{cauchy}) are equal to zero 
 $$
%\begin{eqnarray}\label{cauchy'} 
\nabla^{\perp}_{r'}{\cal H}^{\nu\mu}_{ab}|_{\tau=0}=
\nabla^{\perp}_{r'}\nabla^{\perp[\mu}l^{\nu]\rho a}|_{\tau=0}
=\nabla^{\perp}_{r'}\nabla^{\perp}_{\mu}l^{\mu\rho}_{a}|_{\tau=0}=0, \ \ \ \ (r'=1,2,...,p)
 $$
 derive conservation of the part of desired Ricchi-Codazzi conditions 
\begin{eqnarray}\label{zerosp} 
{\cal H}^{\tau r}_{ab}(\tau,\sigma^r)=0, \ \  \nabla^{\perp[\tau}l^{r]\rho}_{a}(\tau,\sigma^r)=0, 
\ \ \nabla^{\perp}_{\mu}l^{\mu\rho}_{a}(\tau,\sigma^r)=0.
\end{eqnarray} 
Vanishing of the magnetic components  ${\cal H}^{s r}_{ab}$ follows from 
their definition %of ${\cal H}^{\nu\mu}_{ab}$
 and the first pair of the Maxwell equations for $H^{\mu\nu}_{ab}$ 
 resulting in the equations
\begin{eqnarray}\label{Maxw'} 
\sum_{cycle(\mu\nu\rho)}(\nabla^{\perp}_{\mu}{\cal H}_{\nu\rho}^{ab}+
\nabla^{\perp}_{[\mu}l_{\nu]\lambda}^{[a}l^{b]\lambda}_{\rho})=0,
\end{eqnarray}  
where ${\sum}_{cycle(\mu\nu\rho)}$ denotes the sum in the cyclic permutations 
of the $\mu,\nu,\rho$ indices.
Eqs. (\ref{Maxw'}) are easily derived using the identities 
\begin{eqnarray}\label{ident}
\sum_{cycle(\mu\nu\rho)}
\nabla^{\perp}_{\mu}(l_{\lambda[\nu}^{a}l^{\lambda b}_{\rho]})=
\sum_{cycle(\mu\nu\rho)}
\nabla^{\perp}_{[\mu}l_{\nu]\lambda}^{[a}l^{b]\lambda}_{\rho}.
\end{eqnarray} 
By combining equations (\ref{Maxw'}), (\ref{zerosp}) with the Cauchy 
data (\ref{cauchy}) we obtain ${\cal H}^{sr}_{ab}(\delta\tau,\sigma^r)=0$. 
%to be the solution of the reduced EOM (\ref{maxH2*}). 
%reduce to the additional equations for $l_{\mu\nu}^{a}$ %\nabla^{\perp[\mu} 
%\begin{eqnarray}\label{maxH2*lL} \nabla^{\perp}l_{\nu\rho[a}\nabla^{\perp[r} l^{\nu]\rho}_{b]} =0.
%\end{eqnarray}
The substitution of all the above found solutions to  (\ref{eqgW2sr}) and
 (\ref{maxH2*}) permits to show that $\nabla^{\perp[r}l^{s]\tau}_{a}(\delta\tau,\sigma^r)
 =\nabla^{\perp[r}l^{s]q}_{a}(\delta\tau,\sigma^r)=0$. 
The latter proves conservation of the complete set of the Codazzi conditions 
$\nabla^{\perp[\mu}l^{\nu]\rho}_{a}=0.$ 
  Thus,  the R-C eqs. (\ref{Hlsol}) are actually conserved in time
\begin{eqnarray}\label{Hlsol'} 
{\cal H}_{\mu\nu}^{ab}(\delta\tau,\sigma^r) ={\cal H}_{\mu\nu}^{ab}(0,\sigma^r ), \ \ \ \ \  
\nabla_{[\mu}^{\perp}l_{\nu]\rho}^{a}(\delta\tau,\sigma^r)
=\nabla_{[\mu}^{\perp}l_{\nu]\rho}^{ a}(0,\sigma^r)
\end{eqnarray} 
together with their consequences $\nabla^{\perp}_{\mu}l^{\mu\rho}_{a}(\tau,\sigma^r)=0$
following from the minimality conditions (\ref{minco}).
In correspondence with the Cauchy-Kowalevskaya theorem of local 
existence and uniqueness, we see that Eqs. (\ref{Hlsol}) 
define the covariant solution of EOM (\ref{maxH2*}-\ref{eqgW2*}) 
modulo the gauge and diffeomorphism symmetries of $S_{Dir}$. 
Then these EOM become equivalent to the identities  
\begin{eqnarray}\label{iden}
\nabla^{\perp\mu}{\cal H}_{\mu\nu}^{ab}=0, \ \ \ \
\nabla^{\perp\mu} \nabla^{\perp}_{[\mu}l_{\nu]\rho}^{a}=0
\end{eqnarray} 
produced by the covariant differentiation 
of the Ricci-Codazzi Eqs. (\ref{Hlsol}),  
and can be equivalently written in the form of the generalized 
Maxwell-Y-M and Newton equations in the gravitational field  
defined by Eqs. (\ref{cRl})
\begin{eqnarray}
\nabla^{\perp}_{\nu}H^{\nu\mu}_{ab}= j^{\mu}_{ab},  \ \ \ \ \ % 
j^{\mu}_{ab}= Sp(l_{[a}\nabla^{\perp\mu}l_{b]}), \ \ \ 
\nabla^{\perp}_{\mu}j^{\mu}_{ab}=0,
 \label{gMax}\\ 
\nabla^{\perp}_{\mu} \nabla^{\perp\mu}l^{\nu\rho a}
=\frac{1}{2} \frac{\partial {V_{Dir}}}{\partial l_{\nu\rho a}}
\equiv 
(2l_{b}l^{a}l^{b}-l^{a}l_{b}l^{b}-l_{b}l^{b}l^{a})^{\nu\rho} 
-l^{\nu\rho}_{b}Sp(l^{b}l^{a}). \label{gNew}
\end{eqnarray}
We conclude that $S_{Dir}$ (\ref{actnl}) 
with the chosen potential $V_{Dir}$ (\ref{solVl})
reformulates the Dirac $p$-brane dynamics it terms of the 
Cartan multiplets.
The particular solution $V=const, \ l_{\mu\nu}^{a}=0$ 
describes flat branes with $g_{\mu\nu}=\eta_{\mu\nu}$.

In the original brane action $S$ (\ref{Dirbr}) 
the N-G translational modes are condenced in the 
volume of the coset $ISO(1,D-1)/SO(1,D-1)$ expressed 
through the derivatives of $\mathbf{x}$. This action
is  the leading term in the long-wave description of  
the brane dynamics. 
In the action $S_{Dir}$ (\ref{actnl}) the translational 
 modes form the background metric $g_{\mu\nu}$ 
 treated as an independent field. 
 The broken translational and rotational modes condenced 
 in $l_{\mu\nu}^{a}$ associate with the extrinsic curvature 
 of $\Sigma_{p+1}$. 
  The Gauss conditions (\ref{cRl}) connect the rotational 
  and translational N-G modes and defines 
 dynamics of the metric field $g_{\mu\nu}$. 
 $S_{Dir}$ also contains the cosmological term  
that points to spontaneous breakdown of the global
 Poincare symmetry of the Minkowski space.

\section {Summary}

The gauge theory approach to branes was interpreted in the language 
used for the system with spontaneously broken internal symmetries.  
However, in contrast to the standard description of the Nambu-Goldstone 
fields as coordnates of a coset $G/H$, 
we considered their purely geometric description %of 
without any explicit parametrization. 
It was based on the use of moving frames, the Cartan forms and the 
 {\it right gauge} symmetries  $H_{R}$.  These symmetries  remove
auxiliary N-G modes corresponding to the generators of 
the vacuum subgroups $H$ of the completely broken global symmetry $G$. 
This shows equivalency of the N-G fields 
 to the Cartan multiplets formed by the constrained 
 vector and Yang-Mills multiplets of $H_{R}$. 
%After clarifying the role of the right gauge symmetry in field theory 
%as a new Lagrangian symmetry additional to broken internal symmetris 
Then we extended this approach to $p$-branes embedded 
into Minkowski space $\mathbf{R}^{1,D-1}$ invariant under 
the global Poincare symmetry $ISO(1,D-1)$. 
The latter was treated as the symmetry spontaneously
broken by the presence of a minimal brane
hypersurface $\Sigma_{p+1}$. 
We treated the orthonormal vectors of the Cartan moving 
frame $\mathbf{n}_{A}(\xi)$ attached 
to $\Sigma_{p+1}$ together with its world vector $\mathbf{x}(\xi)$
 as the order parameters fixing a macroscopic 
 vacuum manifold of p-brane represented by $\Sigma_{p+1}$. 
 The symmetry group of the vacuum manifold was fixed 
by the condition of vanishing for the translational Cartan 
forms $\omega_{a}$ orthogonal to $\Sigma_{p+1}$. 
This resulted in emergence of constrained 
Cartan multiplets of the unbroken subgroup $SO(D-p-1) \in SO(1,D-1)$,
 their %$(p+1)$-dimensional 
gauge invariant interaction potential $V_{Dir}$ (\ref{solVl}) 
and the action $S_{Dir}$ (\ref{actnl}). 
 The multiplet constraints were treated as the conserved 
 Cauchy data for the corresponding Euler-Lagrange EOM.
 Thus, $S_{Dir}$  was shown to give an alternative description   
 of the fundamental $p$-branes by the Yang-Mills (\ref{gMax}) 
 and Newton (\ref{gNew}) equations.
 Thereat, the evolution of p-brane metric was encoded by 
 the Gauss conditions (\ref{cRl}) treated as the second order PDEs.
For co-dimension 1 ${H}_{\mu\nu}^{ab}=j_{\mu}^{ab}\equiv 0$ and 
Eqs. (\ref{gMax}-\ref{gNew}) reduce to the (p+1)-dim. equation 
\begin{eqnarray}\label{Newt}
\Box l_{\nu\rho}
=l_{\nu\rho}Sp(l^2),\ \ \ \ \ Spl=0, 
\end{eqnarray} 
where $\Box\equiv\nabla_{\mu} \nabla^{\mu}$ is the 
D'Alembert-Beltrami operator for a tensor field on $\Sigma_{p+1}$.
Eq.  (\ref{Newt}) coincides with 
 the Laplace-Beltrami
 equation for the second fundamental form 
 of minimal hypersurfaces in Euclidean 
 space (\cite{JS},\cite{SSY}), 
 but with the $\Box $ operator substituted 
 for the Laplace-Beltrami one. 
  \footnote{The author is indebted to G. Huicken for this observation.} 

The gauge approach can be generalized to the case of $D$-branes 
or $M$-theory branes. This will modify 
the potential $V$ in the action (\ref{actn2}). 
It is also interesting to quantize $S_{Dir}$ using  
 the well-known BRST-BFV method. Quantization may fix the
 cosmological constant value and shed new light on the problems 
 connected with ghosts, anomalies and critical dimensions.

\vskip 10 pt

\noindent{\bf Acknowledgments}
\vskip 10 pt

The author would like to express his thanks 
to NORDITA and Physics Department of Stockholm University  
 for kind hospitality and support, to 
  J. Buchbinder, A. Rosly, H. von Zur-Muhlen for stimulating discussions, 
to S. Krivonos who pointed to reference \cite{BKS}. 
I would also like to thank G. Huicken for very interesting
discussion and references \cite{JS} and \cite{SSY}.


\begin{thebibliography}{99}

\bibitem{RL}
F. Lund and T. Regge, 
Unified approach to strings and vortices with soliton solutions,
Phys. Rev. D 14 (1976) 1524-1535. 

\bibitem{Om}
R. Omnes, A new geometric approach to the relativistic string, 
 Nucl. Phys. B 149 (1979) 269-284.

\bibitem{BNCh}
B.M. Barbashov, V.V. Nesterenko and A.M. Chervyakov, 
On the theory of world surfaces of a constant mean curvature,
Theor. Math. Phys. 21 (1979) 15-41.

\bibitem{BN}
B.M. Barbashov and V.V. Nesterenko, Introduction to the Relativistic 
String Theory (World Scientific Pub Co Inc, Singapore, 1990).

\bibitem{Eisn}
L.F. Eisenhart,
Riemannian Geometry (Princeton University Press, Princeton, 1968).

\bibitem{AKNS}
M. Ablowitz, D.J. Kaup, A.C. Newel and H. Segur, 
Method for solving the sine-gordon equation,
 Phys. Lett. 30 (1973) 1262-1264.

\bibitem{hoppe1}
J. Hoppe, Quantum theory of a massless relativistic surface,
in Proc. Int. Workshop on Constraints 
Theory and Relativistic Dynamics, eds. by G. Longhi and L. Lusanna 
(World Scientific, Singapore, 1987), pp. 267-276; 
MIT PhD Thesis (1982).
 
\bibitem{BST}
 E. Bergshoeff, E. Sezgin and P.K. Townsend, 
 Supermembranes and eleven-dimensional supergravity,
 Phys. Lett. B 189 (1987) 75-78. 

\bibitem{DHIS}
M. Duff, P. Howe, T. Inami and K. Stelle, Superstrings in D = 10 
from supermembranes in D = 11,
Phys. Lett. B 191 (1987) 70-74.

\bibitem{Nic}
 J. Hoppe and H. Nicolai, Relativistic minimal surfaces,
 Phys. Lett. B 196 (1987) 451-455.

\bibitem{WHN}
 B. de Witt, J. Hoppe and G. Nicolai, On the quantum mechanics
  of supermembranes, Nucl. Phys. B 305 (1988) 545-581.

\bibitem{FI} 
 E. Floratos and J. Illipoulos, A note on the classical
  symmetries of the closed bosonic membranes, 
  Phys. Lett. B 201 (1988) 237-240.

\bibitem{WLN}
 B. de Witt, M. Lusher and G. Nicolai, The supermembrane is 
 unstable, Nucl. Phys. B 320 (1989) 135-159.

\bibitem{BZ_0}
I.A. Bandos and A.A. Zheltukhin, Null super p-brane quantum theory in 
4-dimensional space-time, Fortschr. Phys. 4 (1993) 619-676. 

\bibitem{Twn}
P. K. Townsend, The eleven-dimensional supermembrane revisited, 
Phys. Lett. B 350 (1995) 184-188. 

\bibitem{Wit}
 E. Witten, String theory dynamics in various dimensions, 
 Nucl. Phys. B 443 (1995) 85-126.

 \bibitem{Duff}
M. J. Duff, The World in Eleven Dimensions: Supergravity, Supermembranes 
and M-theory (IOP, Bristol, 1999).

\bibitem{TZ}
A.A. Zheltukhin and M. Trzetrzelewski, U(1)-invariant membranes:
 The geometric formulation, Abel, and pendulum differential equations,
 J. Math. Phys. 51 (2010) 062303. 
%Exact solutions for U(1) globally invariant membranes
%Phys. Lett. B, 679 (2009), pp. 523-28

\bibitem{Hop}
J. Hoppe, U(1)-invariant minimal hypersurfaces in $R^{1,3}$,
Phys. Lett. B 736 (2014) 465-469.

\bibitem{Zgau2}
A.A. Zheltukhin, 
Classical relativistic string as an exactly solvable sector 
of SO(1,1)xSO(2) gauge model,
 Phys. Lett. B 116 (1982) 147-150; A.A. Zheltukhin, 
On relation between a relativistic string and two-dimensional 
field models, Sov. J. Nucl. Phys. 34 (1981) 311-322. 
%, Yad.Fiz. 33 (1981) 1723-1728

\bibitem{Zgau4}
A.A. Zheltukhin,
Gauge description and nonlinear string equations in 
d-dimensional space-time,
Theor. Math. Phys. 56 (1983) 785-795.
% DOI: 10.1007/BF01016820

\bibitem{Car}
E. Cartan, Riemannian Geometry in an Orthogonal Frame 
(World Scientific, Singapore, 2001).

\bibitem{Vol1}
D.V. Volkov,  Phenomenological lagrangian of interaction for goldstone particles.
Kiev preprint ITF-69-75 (1969); D.V. Volkov, Phenomenological lagrangians, 
Phys. of Elem. Part. At. Nucl. 4 (1973) 3-41.
 
\bibitem{Wei}
S. Weinberg, Dynamical approach to current algebra, 
Phys. Rev. Lett. 18 (1967) 188-191.

\bibitem{Schw}
J. Schwinger, Chiral dynamics, Phys. Lett. B 24 (1967) 473-476.

\bibitem{CWZ}
S. Coleman, J. Wess and B. Zumino, Structure of phenomenological
lagrangians. I, Phys. Rev. 177 (1969)  2239-2247.

\bibitem{CCWZ}
C. Callan,  S. Coleman, J. Wess and B. Zumino, 
Structure of phenomenological lagrangians. II, 
 Phys. Rev. 177 (1969) 2247-2250.

\bibitem{Zbran}
A.A. Zheltukhin, On brane symmetries,
Phys. Part. Nucl. Lett. 11(7) (2014) 899-903. 
%DOI: 10.1134/S1547477114070486 
A.A. Zheltukhin, Branes as solutions of gauge theories 
in gravitational field, Eur. Phys. J. C 74 (2014) 3048 (9 pp.).
%Conference: C13-07-29.6 Proceedings; \\
%e-Print: arXiv:1409.6655 [hep-th] 
 
\bibitem{FST}
M.A. Semenov-Tyan-Shansky, L.D. Faddeev, 
To the theory of nonlinear chiral fields,
Vestnik St. Petersburg Univ.  13(3) (1977) 81-88 (in Russian).

\bibitem{BCGM}
J. Brugues, T. Curtright, J. Gomis and L. Mezincescu,
Non-relativistic strings and branes as non-linear realizations 
of Galilei groups,
Phys. Lett. B 594 (2004) 227-233.

\bibitem{GKW}
J. Gomis, K. Kamimura and P. West,
The construction of brane and superbrane 
 actions using non-linear realizations, 
Class. Quant. Grav. 23 (2006) 7369-7381.

\bibitem{BZ_hedr}
I.A. Bandos and A.A. Zheltukhin, 
Spinor Cartan moving n-hedron, Lorentz harmonic formulations 
of superstrings, and kappa symmetry,
  JETP Lett. 54 (1991) 421-424.

\bibitem{CLNVX}
T.E. Clark, S.T. Love, M. Nitta, T. ter Veldhuis and C. Xiong,
Oscillating $p$-Branes, Phys. Rev. D 76 (2007) 105014. 
%[arXiv: hep-th/0703179].
 
\bibitem{GM}
F. Gliozzi, M. Meineri, Lorentz completion of effective string 
  (and p-brane) action, JHEP 1208 (2012) 056.

\bibitem{AKo}
 O. Aharony and Z. Komargodski, The effective theory of long 
 strings, JHEP 305 (2013) 118.
 % [arXiv:1302.6257[hep-th]].

\bibitem{GKP}
J. Gomis, K. Kamimura and J. M. Pons,
Non-linear realizations, Goldstone bosons of broken Lorentz 
rotations and effective actions for p-branes,
Nucl. Phys. B 871 (2013) 420-451.
%Class. Quant. Grav. 23, 7369 (2006).
%[arXiv: 1205.1385[hep-th]].

\bibitem{VGZP}
D.V. Volkov, V.D. Gershun, A.A. Zheltukhin, A.I. Pashnev, 
Adler principle and algebraic duality,
Theor. Math. Phys. 15 (1973) 495-504.
% ; DOI: 10.1007/BF01028224.
%Published in Teor.Mat.Fiz. 15 (1973) 245-258 
%DOI: 10.1007/BF01028224 % Volume 15, Issue 2, pp 495-504
 
\bibitem{deW} B. S. De Witt,
 Dynamical theory of groups and fields (Gordon and
Breach, New York, 1965).

\bibitem{FV}
 E.S. Fradkin and G.A. Vilkovisky,  Quantization of relativistic 
 systems with constraints,
 Phys. Lett. B 55 (1975) 224-226.

\bibitem{IO} E.A. Ivanov and V.I. Ogievetsky,
The inverse Higgs phenomenon in nonlinear realizations, 
Teor. Mat. Fiz. 25 (1975) 164-177.

\bibitem{VZT}
D.V. Volkov,  A.A. Zheltukhin, V.I. Tkach,
On minimal interaction of $\pi$-mesons,
Theor. Math. Phys. 10 (1972) 218-224.  
%DOI: 10.1007/BF01035667.

\bibitem{Dir} 
P.A.M. Dirac, Long range forces and broken symmetries,
Proc. R. Soc. Lond. A 333 (1973) 403-418.

\bibitem{BKS}
S. Bellucci, S. Krivonos, A. Sutulin,
Coset approach to the partial breaking of global supersymmetry,
 arXiv:hep-th/1401.2613.

\bibitem{JS}
J. Simons,
Minimal varieties in riemannian manifolds,
Annals of Mathematics, 88 (1968) 62-105.

\bibitem{SSY}
R. Schoen, L. Simon, S. T. Yau, 
Curvature estimates for minimal hypersurfaces,
Acta Mathematica,  134 (1975) 275-288.



\end{thebibliography}
\end{document}